\documentclass[11pt,a4paper]{article}
\pdfoutput=1
\usepackage{jheppub}

\usepackage{amsmath,amssymb}
\usepackage{graphics}

\DeclareMathOperator{\Li}{Li}
\newcommand{\tmop}[1]{\ensuremath{\operatorname{#1}}}
\newcommand{\tmtexttt}[1]{{\ttfamily{#1}}}

\newcommand\fb{{f_b}}
\newcommand\Kinn{{\bf \Phi}_n}
\def\lq{\left[} 
\def\rq{\right]} 
\def\rg{\right\}} 
\def\lg{\left\{} 
\def\({\left(} 
\def\){\right)} 

\newcommand\sss{\mathchoice%
{\displaystyle}%
{\scriptstyle}%
{\scriptscriptstyle}%
{\scriptscriptstyle}%
}
\newcommand\frindsing{{\ctindr}}
\newcommand\ctindr{{\alpha_{\sss\rm r}}}
\newcommand\Rad{\Phi_{\rm rad}}
\newcommand\Kinnpo{{\bf \Phi}_{n+1}}
\newcommand\BKinn{{\bf \bar{\Phi}}_n}
\newcommand\ctindp{{\alpha_{\splus}}}
\newcommand\ctindm{{\alpha_{\sminus}}}
\newcommand\Kinncp{{\bf \Phi}_{n,\splus}}
\newcommand\Kinncm{{\bf \Phi}_{n,\sminus}}
\newcommand\splus{{\sss \nplus}}
\newcommand\sminus{{\sss \nminus}}
\newcommand\nplus{\oplus}
\newcommand\nminus{\ominus}
\newcommand\pt{p_{\sss\rm T}}
\newcommand\kt{k_{\sss\rm T}}
\newcommand\stepf{\theta}
\newcommand\ptmin{{\pt^{\min}}}

\newcommand\POWHEGBOX{\texttt{POWHEG BOX}}
\newcommand\POWHEG{\texttt{POWHEG}}

\newcommand\aem{\alpha_{\rm em}}

\title{Implementation of electroweak corrections in the \POWHEGBOX{}: 
single $W$ production}
\author[a]{\large Luca Barz\`e,}
\author[a]{\large Guido Montagna,}
\author[b,c]{\large Paolo Nason,}
\author[d]{\large Oreste Nicrosini}
\author[d]{\large and Fulvio Piccinini}
\affiliation[a]{Dipartimento di Fisica Nucleare e Teorica, Universit\`a di Pavia
and INFN, Sezione di Pavia,\\
Via A. Bassi 6, 27100 Pavia, Italy}
\affiliation[b]{Theory Division, CERN, CH--1211, Geneva 23, Switzerland}
\affiliation[c]{INFN and Department of Physics, University of Milano Bicocca, 20133 Milan, Italy}
\affiliation[d]{INFN, Sezione di Pavia,Via A. Bassi 6, 27100 Pavia, Italy}
\emailAdd{luca.barze@pv.infn.it}
\emailAdd{guido.montagna@pv.infn.it}
\emailAdd{paolo.nason@mib.infn.it}
\emailAdd{oreste.nicrosini@pv.infn.it}
\emailAdd{fulvio.piccinini@pv.infn.it}
\abstract{We present a fully consistent implementation of electroweak and strong radiative corrections 
to single $W$ hadroproduction
in the \POWHEGBOX{} framework, treating soft and collinear photon emissions
on the same ground as coloured parton emissions. This framework can be easily
extended to more complex electroweak processes.
We describe how next--to--leading order (NLO)  electroweak corrections are combined with the NLO
QCD calculation, and show how they are interfaced to QCD and QED shower Monte Carlo. 
The resulting tool fills a gap in the literature and allows to 
study comprehensively the interplay of QCD and 
electroweak effects to $W$ production using a single 
computational framework.
Numerical comparisons with the predictions of 
the electroweak generator \texttt{HORACE}, as well as with existing results on the combination of electroweak and QCD corrections 
to $W$ production, are shown for the LHC energies, to validate the reliability and accuracy of the approach.}

\begin{document}
\rightline{CERN--PH--TH--2012--025}
\rightline{FNT/2012/01}
\rightline{LPN12--031}
\maketitle

\section{Introduction}

The production of a high transverse momentum lepton--neutrino pair in hadronic collisions, a process known as 
charged current (CC) Drell--Yan (DY), represents one of the most relevant processes at the LHC because of its 
large cross section and clean signature. This reaction is very well suited for $i$) a 
precise determination of two fundamental parameters of the Standard Model (SM), {\em i.e.} the $W$--boson 
mass $M_W$ and width $\Gamma_W$; $ii$) constraining the parton distribution function 
(PDF) of the proton; $iii$) background studies in the search for new heavy resonances;
$iv$) a possible determination of the collider luminosity.

Specifically, thanks to the very large statistics, the LHC might measure $M_W$ with an 
accuracy of about 15 MeV or possibly better, even if this measurement appears 
at present particularly challenging. More generally, the very large statistics 
accumulated at the LHC implies that the systematic errors, 
including the theoretical ones, play a dominant role in the determination of the 
measurement error of the total cross sections, as well as of the other observables of 
experimental interest.

The luminosity and the PDFs are two of the main sources of systematic error for the LHC 
measurements. As an alternative to ordinary measurements of the PDFs, the process of single $W$ 
production could provide a handle to 
constrain the PDFs themselves through the measurement of the $W^+/W^-$ asymmetry, while 
the standard luminosity determinations could be cross--checked with the 
measurement of the total cross section of single vector boson (in particular $Z$) 
production.

Last but not least, in the high energy regions of the phase space, where 
the searches for new physics beyond the SM are focused, the CC DY 
is the main background to unravel signatures due to the presence of heavy 
charged bosons predicted by many extensions of the SM. In these high energy tails, 
the electroweak (EW) loop diagrams
containing the exchange of massive gauge bosons give rise to
large contributions to the experimental observables, because of the presence of
EW Sudakov logarithms $\propto \log^2 (\hat{s}/M_V^2)$, 
$\log (\hat{s}/M_V^2)$, where
$\hat{s}$ is the squared partonic center of mass (c.m.) energy and $M_V, V = W, Z$ the gauge boson mass.

Because of the above reasons, it is essential for the LHC physics programme to have accurate 
theoretical predictions available for the $W$ production cross section and associated distributions. 

In perturbative QCD (pQCD), leading order (LO) calculations have in general a large 
uncertainties due to the dependence upon the unphysical factorization 
and renormalization scales. 
These uncertainties can be reduced including at least pQCD next--to--leading/next--to--next-to--leading order 
(NLO/NNLO) corrections. The Drell--Yan process was one of the first to be 
calculated in  pQCD at NLO accuracy \cite{Altarelli:1979ub}, 
while NNLO predictions for the integrated cross section became available
later in Ref.~\cite{Hamberg:1990np}.
The NNLO pQCD computation of DY observables in a fully differential form
was completed only in recent years by two 
independent groups~\cite{Anastasiou:2003yy,melnikov:2006kv, Catani:2009sm}.

The size of the NNLO pQCD contributions is naively comparable to the one of the NLO EW 
radiative corrections, fully computed in Refs.~\cite{Dittmaier:2001ay,Zykunov:2006yb,Baur:2004ig,
Arbuzov:2005dd,CarloniCalame:2006zq}. The EW corrections are therefore a 
necessary ingredient in a precise theoretical description of the DY observables. 
For example, it is known that they induce a shift on $M_W$ of the order of $100$ MeV~\cite{CarloniCalame:2003ux}
and introduce a negative correction to the transverse mass distribution $M_T^W$ of about 20--30\% for
$M_T^W$ above 1 TeV~\cite{CarloniCalame:2006zq}.

In particular, realistic phenomenological studies and data analysis require the simulation of all the 
experimental cuts and of the detector acceptance. Moreover the 
broad nature of the $W$ physics programme at the LHC relies upon the measurement of 
a number of observables, which must be precisely predicted to avoid any theoretical bias in data interpretation. Hence, the 
implementation of the most important theoretical ingredients into a  Monte Carlo (MC)
event generator is mandatory. The presence of soft and collinear divergences due to the emission 
of coloured particles by the initial state partons leads any fixed--order calculation to provide accurate predictions only for 
quantities sufficiently inclusive over QCD radiation, like the total cross section, the $W$ rapidity and 
transverse mass distribution. However, exclusive quantities, like the $W$ or lepton transverse momentum, are often of great experimental interest,
so that Shower Monte Carlo (SMC) programs, like \texttt{HERWIG} \cite{Corcella:2002jc}, 
\texttt{PYTHIA} \cite{Sjostrand:2007gs} or 
\texttt{SHERPA} \cite{Gleisberg:2008ta}, must be used. Beyond the pure Parton Shower approximation, 
the generators \texttt{POWHEG}~\cite{Nason:2004rx,Frixione:2007vw} 
and \texttt{MC@NLO}~\cite{Frixione:2002ik} are widely employed at the LHC because they guarantee a 
better accuracy, being based on a 
consistent matching of NLO QCD corrections with SMC codes.

Till now, EW and QCD corrections have been typically implemented in
event generators separately. Fixed--order QCD programs available for $W$ production
are \texttt{MCFM}~\cite{Giele:1993dj,Campbell:2002tg} at NLO accuracy, and 
 \texttt{FEWZ}~\cite{melnikov:2006kv,Gavin:2012kw}  and \texttt{DYNNLO} \cite{Catani:2009sm} 
 at NNLO. The \texttt{ResBos} code  \cite{Balazs:1997xd}
is based on analytical resummation of all transverse momentum 
logarithms with NNLO accuracy.
As already said, \texttt{MC@NLO} \cite{Frixione:2002ik} and the \POWHEGBOX{} \cite{Alioli:2010xd} 
combine NLO QCD corrections with SMC, albeit according to a 
different methodology. Programs implementing exact NLO EW corrections
are \texttt{DK} \cite{Dittmaier:2001ay},  \texttt{WGRAD} \cite{Baur:2004ig}, \texttt{SANC} \cite{Arbuzov:2005dd}
and \texttt{HORACE} \cite{CarloniCalame:2006zq}. In  \texttt{HORACE}  NLO EW 
corrections are matched with a QED Parton Shower \cite{CarloniCalame:2003ux}, 
while in \texttt{WINHAC} \cite{Placzek:2003zg} the simulation
of multiple photon effects, realized through the YFS framework, 
is associated with NLO EW contributions through an interface to the
\texttt{SANC} module \cite{Bardin:2008fn}. A QED shower for final state particles can be also described by other 
codes, using {\it e.g.} a universal and widely used tool like \texttt{PHOTOS} \cite{Golonka:2005pn}. 

Only  during the last few years the interplay of QCD and EW corrections to the CC DY observables 
has been studied in some detail. In Ref.~\cite{Cao:2004yy} the combined effect of QCD resummation and 
NLO QED final--state corrections to $W$ boson observables has been addressed, while a 
more thorough analysis of the EW$\otimes$QCD interplay has been performed in Ref.~\cite{Balossini:2009sa}, using, among others, 
the  \texttt{HORACE} and  \texttt{MC@NLO} programs. 
First attempts towards a single framework implementation of NLO EW corrections 
in association with QED and QCD showers are documented in 
Ref.~\cite{Richardson:2010gz} and Ref.~\cite{Yost:2012mf}. 
The EW corrections to $W + $jet production (with and without $W$ leptonic decays), which are
part of the ${\cal O}(\aem \alpha_s)$ corrections to inclusive $W$ production, were computed 
in Refs. \cite{Kuhn:2007qc,Kuhn:2007cv,Hollik:2007sq,Denner:2009gj}.
These studies outlined the importance of the simultaneous control of 
all the relevant QCD and EW corrections for a sensible description of $W$ boson observables.

In the light of the above motivations,  it would be desirable to have at our 
disposal
\begin{itemize}
	\item a MC event generator for $W$ production
	\item with NLO QCD and NLO EW corrections
	\item interfaced to QCD and QED SMC.
\end{itemize}
Such a tool is presently unavailable.\footnote{Note in fact that the 
inclusion of EW radiative corrections into a QCD generator as described in 
Refs.~\cite{Richardson:2010gz,Yost:2012mf} 
is limited to the universal, Parton Shower description of QCD radiation, thus neglecting 
NLO QCD corrections.} We have built this program, and the 
aim of the paper is describing the theoretical and technical steps followed to achieve this goal. 
We also present first numerical results to cross--check the correct calculation and codification of
all the different ingredients. Since we are generally interested in 
extending the
present approach to processes other than $W$ production, we have chosen the 
 \POWHEGBOX{} as a general computer framework for implementing our generator.

The rest of the paper is structured as follows. In Section~\ref{sec:powheg} we
shortly review the \POWHEGBOX{} framework, 
and summarize the basic modifications needed for a
consistent inclusion of the EW effects.
In Section~\ref{sec:nlo} the main details of the full NLO EW computation 
are described, as well as its concrete implementation in the \POWHEGBOX{}. 
In Section~\ref{sec:implementation} we sketch the modifications introduced 
in the \POWHEGBOX{} to take care of photon
radiation. In the same Section we also illustrate the interface 
of NLO corrections with a mixed QCD$\otimes$QED shower. 
A sample of numerical results, both for NLO EW corrections and for the full 
combination of QCD and EW contributions, in shown in Section~\ref{sec:results}. 
Conclusions and perspectives are drawn in Section \ref{sec:concl}.

\section{The \POWHEGBOX{}: framework and basic electroweak modifications}
\label{sec:powheg}
The \texttt{POWHEG} method (\texttt{Po}sitive \texttt{W}eight
\texttt{H}ardest \texttt{E}mission \texttt{G}enerator), first
introduced in Ref.~\cite{Nason:2004rx}, has been used so far to implement a
wide variety of NLO QCD corrections into a NLO + Parton Shower matching
framework.\footnote{See \url{powhegbox.mib.infn.it} for a list of
  implemented processes.} It has been described in great detail in
Ref.~\cite{Frixione:2007vw}. Recently the \texttt{POWHEG} algorithm
has been codified into the \POWHEGBOX{}~\cite{Alioli:2010xd},
which is a general computer framework for implementing NLO
calculations in SMC programs according to the \texttt{POWHEG} method.
Using the \POWHEGBOX{}, the implementation of a QCD NLO process requires
the following ingredients:
\begin{itemize}
	\item the squared Born matrix elements $\mathcal{B}$;
	\item the colour correlated Born matrix elements $\mathcal{B}_{ij}$;
	\item the Born phase space;
	\item the squared matrix element $\mathcal{R}$ of the real emission process;
	\item the finite part of the virtual correction, $\mathcal{V}$, computed in dimensional regularization 
		  and divided by the factor $\mathcal{N}=\frac{(4\pi)^\epsilon}{\Gamma(1-\epsilon)}\left(\frac{\mu^2}{Q^2}\right)^\epsilon$. 
		  As specified in the next Section, $\epsilon$ and $\mu$ are the usual parameters of the calculation in dimensional
		  regularization, and $Q^2$ is the squared virtuality scale of the process under consideration.
\end{itemize}
The remaining parts of the NLO calculation is performed by \POWHEGBOX{} itself, that
\begin{itemize}
    \item identifies all the singular regions (corresponding to soft and collinear emission);
    \item projects the real emission contributions over the singular regions;
    \item implements the subtraction procedure;
    \item computes the soft and collinear remnants;
    \item generates the event with the hardest radiation.
\end{itemize}
In order to include EW corrections together with the NLO QCD ones,
several extensions of the \POWHEGBOX{} were needed. In fact,
collinear singularities due to the emission of photons must also be
consistently included, and one should also foresee that the hardest
radiation could be given by a large transverse momentum
photon, rather than a coloured parton.\footnote{We do not
  consider among the real emission processes the
  contributions due to emission of gauge bosons other than the photon,
  because $W/Z$ radiation gives rise to final states experimentally
  distinguishable from the process under study.}
In the process we
are considering, the only relevant QCD singular region is the one of
small transverse momentum gluon radiation, {\it i.e.} is the initial state
radiation (ISR).
We must now also consider the EW singular regions, {\it i.e.} the
possibility that also a photon may be produced at small transverse
momentum by the incoming charged quarks. Furthermore, the final state 
lepton may emit a collinear photon. Thus, while in $W$ production
with only NLO QCD corrections there is only one singular
configuration, including EW corrections we end up with three regions:
ISR gluon emission, ISR photon emission and final state radiation
(FSR) photon emission. 

The
general \texttt{POWHEG} formula when more than one singular region is
present is given by Eqs.~(4.14), (4.16) and (4.17) of 
Ref.~\cite{Frixione:2007vw}, that we report here for convenience
\begin{eqnarray}
\label{eq:POWHEGsigma}
d\sigma&=&\sum_\fb \bar{B}^\fb(\Kinn)\, d \Kinn
 \Bigg\{ \Delta^\fb\!\(\Kinn,\ptmin\)
\nonumber\\
&+&\!\!\!\!\!\!\sum_{\frindsing\in\{\frindsing|\fb\}}\!\!\!\!\!\!
 \frac{\Big[  d\Rad\;\stepf\(\kt-\ptmin\)
\Delta^\fb\!\(\Kinn,\kt\)\, R\(\Kinnpo\)
\Big]_\frindsing^{\BKinn^\frindsing=\Kinn}}{ B^\fb\!\(\Kinn\)}
\Bigg\},\phantom{aa}
\end{eqnarray}
where
\begin{eqnarray}
\label{eq:bbdef}
\bar{B}^\fb(\Kinn) &=& \lq
B\(\Kinn\)+V\(\Kinn\)\rq_\fb
+ \sum_{\frindsing\in\{\frindsing|\fb\}} \int \Big[  d\Rad\,\lg R\(\Kinnpo\) -
 C\(\Kinnpo\)\rg\Big]_\frindsing^{\BKinn^\frindsing=\Kinn}
\nonumber \\
&&+ \sum_{\ctindp\in\{\ctindp|\fb\}} \int \frac{dz}{z}
\,G_\splus^\ctindp\(\Kinncp\)
+ \sum_{\ctindm\in\{\ctindm|\fb\}} \int \frac{dz}{z}
\,G_\sminus^\ctindm\(\Kinncm\) \;,
\end{eqnarray}
and
\begin{equation}
\label{eq:suddef}
\Delta^\fb(\Kinn,\pt)=\exp\lg
-\sum_{\frindsing\in\{\frindsing|\fb\}}
\int \frac{\Big[  d\Rad\, R\(\Kinnpo\)\, \stepf\(\kt(\Kinnpo)-\pt\)
\Big]_\frindsing^{\BKinn^\frindsing=\Kinn}}{ B^\fb\(\Kinn\)}
\rg\;.
\end{equation}
The function $\bar{B}^\fb(\Kinn)$ is the NLO inclusive cross section at fixed
underlying Born flavour $\fb$ and kinematics $\Kinn$.
The real contributions are separated
into terms labelled by the index $\frindsing$.
Each $\frindsing$ denotes a single flavour structure, and a single
singular region. Each term $R_\frindsing$ is singular only in the
singular region denoted by $\frindsing$. In our case, for example,
the real contribution with flavour structure
$d\bar{u}\to (W^-\to e \bar\nu) g$ is singular only in the region
of small transverse momentum of the gluon, {\it  i.e.} in the ISR
region. Thus it corresponds to
a single $\frindsing$. On the other hand, $d\bar{u}\to(W^-\to e \bar\nu)\gamma$
is singular in the  region of small transverse momentum of the photon,
and in the region where the photon transverse momentum with respect to 
(w.r.t.) the electron is small, which is a FSR region.
\POWHEG{} separates the real contribution into
two terms denoted by a different $\frindsing$, one term being singular
only in the ISR, and the other in the FSR region.

The notation $\frindsing\in\{\frindsing|\fb\}$ means all the real
singular contributions that have $\fb$ as underlying Born flavour. The
square brackets with subscript $\frindsing$ and superscript
$\BKinn^\frindsing=\Kinn$ mean that everything inside refers to the
particular real contribution labelled by $\frindsing$, and having
underlying Born kinematics equal to $\Kinn$.

The Sudakov form factor, Eq.~(\ref{eq:suddef}), is a product of
individual Sudakov form factors associated with each $\frindsing$.
The \POWHEGBOX{} is already capable of generating radiation
using this Sudakov form factor, using the highest bid method
(see Ref.~\cite{Frixione:2007vw}). The generation of radiation
must however be improved to allow for the generation of photon
emission.

In summary,
the \POWHEGBOX{} was modified in the following points:
\begin{itemize}
\item
the term $V$ in Eq.~(\ref{eq:bbdef}) is the sum of the soft
and virtual contributions. The soft contribution is evaluated by
the \POWHEGBOX{}, and now it must also include the terms arising
from soft or collinear photons.
\item
The routines that automatically search for the singular regions in the
real contributions have been modified, so that also a photon--quark and
a photon--lepton pair of external lines are recognized as a singular
region.
\item
The routines that compute the soft, collinear and soft--collinear limit
of the real amplitude had also to be modified with the inclusion of the
soft and collinear photon regions.
\item
The routine that computes the collinear remnants must also deal
with photons.
\item
The computation of the EW virtual corrections are done keeping the mass
of the lepton finite. FSR in the original \POWHEGBOX{} considers only
strictly massless particles. In fact, massive quarks in the \POWHEGBOX{}
are treated as really heavy, and not giving rise to collinear divergences.
In the present case, the electron is really too light for this approach
to be sensible. Thus, an extension of the \POWHEGBOX{} to allow for
collinear radiation from massive particles has been set up. This feature
is not specific to EW corrections and can also be adopted for not so
heavy quarks.
\end{itemize}

Notice that we also include the real processes with an incoming
gluon, like $g d\to(W^-\to e\bar\nu)u$. We may wonder whether the
analogous EW process $\gamma d\to(W^-\to e\bar\nu)u$ 
(the so--called photon--induced process) should be included.
Although it would be not difficult to include it, we decided to leave
it out, basically for two reasons. The first reason is that there are not many
PDF sets that include electromagnetic evolution, as further remarked in 
Section~\ref{sec:sub}. The second reason is that
the photon density is already very small, of the order of $\aem$, and it
multiplies a process of the order of $\aem$ in this case. One should in fact
remember that the enhancement of initial state electromagnetic radiation
is a logarithmic factor of $\log E/\lambda$, where $\lambda$ is the
collinear cutoff. While for an electron beam $\lambda$ is the mass
of the electron, in a hadron beam $\lambda$ is of the order of a typical
hadronic scale, {\it i.e.} few hundred MeV, three orders of magnitudes larger
than the electron mass. Thus, the one order of magnitude enhancement
of ISR in processes with incoming leptons is reduced to a factor of a few
for hadron beams. These arguments are supported by the numerical 
evaluation of photon--induced processes given in Refs.~\cite{Arbuzov:2007kp,Brensing:2007qm}.

In order to be compliant with the general NLO QCD structure of virtual
corrections in the \POWHEGBOX{}, it is convenient to calculate
also the virtual EW contributions in dimensional regularization of
infrared divergences, which is not the scheme under which perturbative
EW calculations are usually carried out in the literature. Moreover, such an
approach simplifies, at least in principle, the extension of EW
corrections to more complicated processes through the adoption of
available automated procedures. The calculation of the EW virtual
corrections in dimensional regularization is described in Section
\ref{sec:nlo}.

To summarize, the newly developed tool
\begin{itemize}

\item ensures normalization with NLO QCD + EW accuracy

\item combines the complete SM NLO corrections with a mixed
  QCD$\otimes$QED parton cascade, where the particles present in the
  shower are coloured particles or photons

\item consequently, incorporates mixed ${\cal O}(\aem \alpha_s)$ contributions
  with a better accuracy w.r.t. existing programs. In particular, it
  can allow to study consistently the interplay between QCD and EW
  radiation, like {\it e.g.} the link between a photon emitted after QCD
  radiation and viceversa.

\end{itemize}

\section{NLO electroweak calculation}
\label{sec:nlo}

In the present section we describe the main features of the NLO EW
calculation. We first detail the treatment of the virtual
contributions, then we discuss finite--width effects, and finally we
discuss the subtraction procedure of infrared and collinear
divergencies.
\subsection{Virtual contributions in dimensional regularization}
The typical framework used in NLO EW calculations to regularize
infrared (IR) and collinear singularities in the loop integrals is the
procedure known as \emph{mass regularization}. The IR soft singularity
is regularized by introducing an infinitesimal photon mass $\lambda$.
The mass of the lepton acts as physical regulator of the 
lepton--photon collinear singularities.
If quarks take part to the process under consideration, small quark
masses are introduced as regulators of the associated collinear
singularities.\footnote{We observe that these masses have no physical
meaning, {\it i.e.} cannot be considered the physical regulator
of the singularity. In fact, for quark virtualities that approach
typical hadronic scale, perturbation theory breaks down and other
complex physical mechanisms act as regulators of soft singularities.}
The IR divergences and part of the quark mass
singularities cancel in the combination of virtual and real
corrections, while the remaining collinear singularities, due to
photon radiation off the initial state quarks, are absorbed into
the PDFs. Large logarithms of the form $\log(\hat{s}/m_l^2)$, with
$m_l$ lepton mass, originated from the emission of collinear photons
from leptons, as those coming from the decays of the $W/Z$ bosons in
the DY process, survive in not fully-inclusive observables.
Ultraviolet (UV) divergences are universally treated in
\emph{dimensional regularization}.

In QCD calculations dimensional regularization is used to regularize
UV, IR and collinear divergencies. This scheme is implemented in the
\POWHEGBOX{} as well. In order to make the
treatment of soft and collinear singularities uniform in the QCD and
EW sector, we treated the IR and collinear divergencies of EW nature
according to the dimensional regularization as well.  More precisely,
we use a sort of hybrid scheme, adopting dimensional regularization
for the singularities associated to the coloured charged particles and
the photon, but keeping the mass of the leptons finite, since this
mass has a well--defined physical meaning, {\it i.e.} it is the true physical
regulator of QED mass singularities.

In practice, the squared matrix element (ME) for the CC DY process $q
q'\rightarrow l \nu_l$, including one--loop QCD and EW 
virtual corrections, can
be written as:
\begin{equation}
|\mathcal{M}_{QCD+EW}^{\rm one \, loop}|^2 = (1 + 2 \,
\Re\{\delta_{QCD}\} + 2 \, \Re\{\delta_{EW}\})|\mathcal{M}_0|^2 \, , 
\label{eq:oneloop}	
\end{equation}
where $\mathcal{M}_0$ is the ME in the LO approximation and
$\mathcal{M}^{QCD}_1 \equiv \delta_{QCD}\mathcal{M}_0$ is the ME with
NLO QCD virtual corrections, calculated in Ref.~\cite{Altarelli:1979ub} and
already present in the available version of the \POWHEGBOX{}. In Eq.~(\ref{eq:oneloop}) $\mathcal{M}^{EW}_1
\equiv \delta_{EW}\mathcal{M}_0$ is the novel ingredient, 
{\it i.e.} the NLO
EW one--loop ME. Instead of doing {\it ab initio} a new complete
calculation of the EW virtual corrections, we chose, for simplicity,
to rely upon the one--loop structure yielding $\delta_{EW}$ as given
in Ref.~\cite{Dittmaier:2001ay}.  The latter calculation was indeed
cross--checked over the years against the predictions of various
independent EW codes and found in perfect agreement.
According to Ref.~\cite{Dittmaier:2001ay}, $\delta_{EW}$ is a linear
combination of the fundamental 't Hooft--Veltman scalar functions
$B_0$, $C_0$ and $D_0$ \cite{'tHooft:1978xw} and relevant
derivatives. Therefore, their expressions in dimensional
regularization can be obtained by calculating each divergent scalar
function in $n = 4-2\epsilon$ dimensions.  In particular, we
calculated the $B_0$ functions for different energy scale and mass
combinations, while for the $C_0$ and $D_0$ we referred 
to Ref.~\cite{Dittmaier:2003bc} and to Ref.~\cite{Denner:2010tr}, 
respectively. In particular we resorted to Eqs.~(B.2), (B.4), (B.12) and (B.16) from 
Ref.~\cite{Dittmaier:2003bc} and used Eq.~(3.78) and Eq.~(4.19) from 
Ref.~\cite{Denner:2010tr}.  For consistency with the FKS \cite{Frixione:1995ms}
subtraction procedure already present the \POWHEGBOX{}, each scalar
function was divided by the overall factor \cite{Alioli:2010xd}
\begin{equation}
\mathcal{N}=\frac{(4\pi)^\epsilon}{\Gamma(1-\epsilon)}
\left(\frac{\mu_{EW}^2}{Q_{EW}^2}\right)^\epsilon \, ,
\end{equation}
where $\mu_{EW}$ is the (arbitrary) mass scale of dimensional
regularization and $Q_{EW}$ the squared energy scale of the process.\footnote{We checked 
that the implemented library for 
the calculation of the scalar form factors agrees with the output 
of {\tt LoopTools}~\cite{Hahn:2000jm} for the regular scalar functions and with that of 
{\tt GOLEM}~\cite{Cullen:2011kv,Binoth:2008uq} for the 
IR ones.}

Before matching the virtual corrections with the standard (FKS)
subtraction procedure of the \POWHEGBOX{}, we verified, as an
intermediate but not trivial check, that all the $1/\epsilon^2$ and
$1/\epsilon$ poles, as well as the arbitrary scale $\mu_{EW}$, cancel
out up to numerical precision when adding to the virtual part massive
\cite{Catani:2002hc} dipole subtraction formulae \cite{Catani:1996jh} 
improved by the inclusion of the singular collinear contributions from 
the PDFs. Also cancellation of the UV divergences was successfully checked.

We performed further stringent internal cross checks, in order to test also
the finite part of the loop corrections. In particular, in order to
check the accuracy of our implementation of $\delta_{EW}$ in the
\POWHEGBOX{} program, we calculated the basic scalar functions in the
mass regularization scheme and compared, point by point of the phase
space, the corresponding numerical values of the virtual part of
$\delta_{EW}$ with the ones returned by the \texttt{HORACE} code,
which relies upon the same mass regularization scheme.  We observed
that the two programs perfectly agree, at the level of one part over
10$^8$ everywhere in the phase space with the only exception of the
``low" (below $\sim$ 10 GeV) invariant mass region. This disagreement
can be easily understood because \texttt{HORACE} includes part of the
finite mass terms in the virtual correction, while $\delta_{EW}$ does
not.  However, the observed small discrepancy can be considered
acceptable, the low invariant mass region giving a negligible
contribution to the total cross section and being practically
unimportant for physics studies at the LHC. The numerical comparisons
between \texttt{HORACE} and the \POWHEGBOX{} at the level of full NLO
EW corrections shown in Section \ref{sec:results} support this
expectation.

Concerning renormalization and the choice of the EW input parameters,
it is known that a particularly convenient scheme for the calculation
of the EW corrections to the CC DY is provided by the so--called
$G_\mu$ scheme, wherein $G_\mu$ (the Fermi constant), $M_W$ (the $W$
mass) and $M_Z$ (the $Z$ mass) play the role of primary quantities.
Actually, this choice allows to remove the dependence of the results
on the masses of the light quarks entering the self--energy loop
diagrams and responsible for the transition from the fine--structure
constant $\aem(0)$ to the running electromagnetic coupling
$\aem(Q^2)$ at a high--energy scale $Q^2$, {\it e.g.} $Q^2 = M_Z^2$. In the
$G_\mu$ scheme the weak coupling constant $g$ is calculated in terms
of $G_\mu$ and $\sin\theta_W$ instead of $\aem(0)$ and
$\sin\theta_W$.  Consequently, the $G_\mu$ parametrization of the Born
cross section minimizes the EW correction, since the universal
corrections induced by the running of $\aem$ (as well as by the
$\rho$ parameter) are absorbed in the LO amplitude. However, in the
code we leave to the user the possibility of choosing between the
$G_\mu$ and $\aem(0)$ scheme with the flag \texttt{scheme} in the
input file.\footnote{In the $\aem(0)$ scheme, wherein the input
  parameters are $\aem(0)$, $M_W$ and $M_Z$, the masses of the light
  quarks (kept at a finite value only in the gauge invariant subset of
  fermionic corrections) are chosen in such a way to reproduce the
  hadronic contribution to the photon vacuum polarization.} 
  Obviously, for photon radiation from the charged particles we use the
  Thomson value $\aem = \aem(0)$.
  
\subsection{Finite--width effects}
The relevant regions for DY physics are the regions around the $W$
resonance and the very high energy tails of the distributions above
the TeV scale. In particular, around the $W$ peak the treatment of the
EW corrections requires particular care because of the importance of
this region for precision measurements and the presence of particular
logarithms in the electroweak factors.
Indeed, $\delta_{EW}$ contains logarithms of the form $\log(s - M^2_W
+ i\epsilon)$ that are singular if $s = M_W^2$.\footnote{We consider
  here for brevity of notation $s$ like the nominal squared
  c.m. energy entering the parton--level EW calculation before
  convolution with the PDFs. It is understood that in the actual
  hadron--level calculation $s$ is replaced by $\hat{s} = x_1 x_2
  s$.} Since these singularities are cured by a Dyson resummation of
the $W$ self--energy loop diagrams, what is usually done in most EW
calculations is substituting each $\log(s - M^2_W + i\epsilon)$ with
the expression $\log(s-M^2_W + i\Gamma_W M_W)$.  This replacement can
be properly applied in the one--loop virtual amplitude, because the
coefficient of the $\log(s - M^2_W)$ term is gauge invariant. We name
this typically adopted procedure \cite{Dittmaier:2001ay,CarloniCalame:2006zq}~\emph{Changing Logarithm Argument}
(CLA).

However, as widely discussed in the literature, the proper description of finite--width effects related 
to a resonance is a delicate point in perturbation theory and, in particular, in NLO EW calculations,
that must provide gauge--invariant results.

To resolve this kind of problems at the level of one--loop calculations, in recent 
years a new, theoretically consistent scheme, known as \emph{Complex Mass Scheme} (CMS) has been 
developed \cite{Denner:2005fg,Denner:2006ic}. With CMS we have 
at our disposal a universal and easy to implement framework to describe finite--width effects
in the phase space regions near and far--off the resonances. Gauge invariance, as
well as all the Ward identities, are preserved in the CMS procedure, 
while unitarity is guaranteed up to $\mathcal{O}(\aem^2)$. A drawback of CMS 
is that EW spurious corrections at the NNLO level are generated, but they are beyond the
required NLO accuracy. In particular, the calculation of the two points scalar functions entering 
the renormalization constants in CMS 
implies their evaluation in terms of a complex momentum $M_W^2 - i\Gamma_W M_W$. This demands an analytic
continuation in the momentum variable to the unphysical Riemann sheet. We have avoided this 
difficulty as suggested in Ref.~\cite{Denner:2005fg}, 
{\it i.e.} by expanding the self-energies appearing in the 
renormalization constants about real arguments. One--loop accuracy is conserved.

In our calculation we decided to leave the possibility to the user 
of choosing between CMS and CLA schemes (the option \texttt{complexmasses} in the input file)
for the calculation of EW corrections.

\subsection{Subtraction procedure and remnants}
\label{sec:sub}

To complete the NLO EW calculation, we calculated the real
photon/gluon emission processes $qq'\rightarrow l \nu_l \gamma$ and
$qq'\rightarrow l \nu_l g$ for massive leptons $l$, to match the phase
space generation with massive leptons described in Appendix~\ref{sec:fsr}.  We checked that our
calculation of the radiative processes agrees numerically with the
output of \texttt{MADGRAPH} \cite{Alwall:2011uj}.

The \POWHEGBOX{} implements automatically the FKS subtraction procedure,
and its extension to EW processes is straightforward. As already emphasized, the routines that
compute the soft and collinear limits of the real amplitude were
updated in order to deal also with photons.

The soft--virtual amplitude, described in detail in Section 2.4.2 of Ref.~\cite{Frixione:2007vw} in the case of massless coloured partons, and
in Section 4.3 of Ref.~\cite{Alioli:2010xd} in the case of massive
partons, were extended to include also the emission
of a photon by a charged particle.  These formulae are easily
obtained from those relative to a gluon emitted by a massive or massless
quark, that we report here for completeness.

Following the same notation of formula (2.99) in Ref.~\cite{Alioli:2010xd},
the EW virtual part can be written as the sum of a singular and 
finite contribution, {\it i.e.}
\begin{equation}
	\mathcal{V}_{EW}=\frac{\aem}{2\pi} \Big((\mathcal{Q}_{EW} +
        \mathcal{I}_{EW})\mathcal{B} + \mathcal{V}_{EW_{fin}}\Big) \, ,
								\label{eq:sv}
\end{equation}
where $\mathcal{B}_{ij}=\mathcal{B}$ for every $i,j$ colour indices,
because the colour structure is not relevant in the EW calculation, 
and $\mathcal{V}_{EW_{fin}}$ stands for the finite part.  In
Eq. (\ref{eq:sv}) $\mathcal{Q}_{EW}$ and $\mathcal{I}_{EW}$ depend on
the masses, momenta and charges of external particles and are given by
(see Eq.~(A.52) of Ref.~\cite{Alioli:2010xd})
\begin{equation}
\begin{array}{ccl}
	\mathcal{Q}_{EW} & = & -\displaystyle\sum_i q_i^2
        \left(\log\displaystyle\frac{\xi_C^2s}{2Q_{EW}^2}-
        \displaystyle\frac{1}{\beta_i}
        \log\frac{1+\beta_i}{1-\beta_i}\right)\\ & &
        -\log\displaystyle\frac{\mu_F^2}{Q_{EW}^2}
        \left[q_{f_{\oplus}}^2 (\frac{3}{2} + 2\log\xi_C) +
          q_{f_{\ominus}}^2 (\frac{3}{2} + 2\log\xi_C) \right] \, .
          \label{eq:above}
\end{array}
\end{equation}
In Eq. (\ref{eq:above}) $q_{f_{\oplus}}$ ($q_{f_\ominus}$) denotes the charge of the $\oplus$ ($\ominus$) incoming 
particle, $\sum_i$ is intended as a sum over the final state charged particles, $q_i$ is the 
charge of the $i$ final state particle and $\beta_i \equiv |\vec{p}_i|/p_{i0}$. 
$\xi_C$ is an arbitrary parameter which is set equal to one in the code, and $\mu_F$ is the 
factorization scale.
The factor 
$\mathcal{I}_{EW}$ can be written as a sum of three terms, denoted as $\mathcal{H}$ (Eq.~(2.101) of 
Ref.~\cite{Frixione:2007vw}), $\mathcal{J}$ (Eq.~(A.28) of Ref.~\cite{Alioli:2010xd}) and $\mathcal{K}$ 
(Eq.~(A.40) of Ref.~\cite{Alioli:2010xd}) related  to a massless--massless, massless--massive and massive--massive charged particle pair, 
respectively.
The $\mathcal{H}$ contribution is given explicitly by
\begin{equation}
\begin{array}{ccccc}
	\mathcal{H} & = & - \displaystyle\sum_{i,j} q_i q_j \sigma_i\sigma_j\Big(&\displaystyle\frac{1}{2}\log^2\displaystyle\frac{\xi_C^2s}{Q_{EW}^2}
															+\log\displaystyle\frac{\xi_C^2s}{Q_{EW}^2}\log\displaystyle\frac{k_i\cdot k_j}{2E_i E_j}
															-\Li_2\displaystyle\frac{k_i\cdot k_j}{2E_i E_j}&\\
			    &   &               & +\displaystyle\frac{1}{2}\log^2\displaystyle\frac{k_i\cdot k_j}{2E_i E_j} - \log\big(1 - \displaystyle\frac{k_i\cdot k_j}{2E_i E_j} 
			                               \big) \log\displaystyle\frac{k_i\cdot k_j}{2E_i E_j} &\Big) \, ,
\end{array}
\end{equation}
where the sum over $i$, $j$ is  a sum over all the pairs of charged
massless particles. The symbol $\sigma_i$ is defined as in Ref.~\cite{Dittmaier:1999mb}: $\sigma_i = +1$ 
for incoming fermions and outgoing anti--fermions, and $\sigma_i = -1$ for outgoing 
fermions and incoming anti--fermions. The $\mathcal{J}$ term reads
\begin{equation}
	\mathcal{J} = -\displaystyle\frac{1}{2}\sum_{m,l} q_m q_l \sigma_m\sigma_l\left[ \log^2\displaystyle\frac{Q_{EW}^2}{s\xi_C^2} - \displaystyle\frac{\pi^2}{6}
															- I_0(k_l,k_m) \log\displaystyle\frac{Q_{EW}^2}{s\xi^2_C} + I_\epsilon(k_l,k_m) \right] \, ,
\end{equation}
where the sum over $m$, $l$ means a sum over all the massive--massless 
particles pairs, ($k_m^2\neq 0$, $k_l^2=0$). $I_0$ and $I_\epsilon$ are given 
by Eq. (A.23) and Eq.~(A.24) of Ref.~\cite{Alioli:2010xd}, respectively. Last, the $\mathcal{K}$ factor is given by 
\begin{equation}
	\mathcal{K} = -\displaystyle\frac{1}{2}\sum_{m,n} q_m q_n \sigma_m\sigma_n\left[ - I_0(k_m,k_n)\log\frac{Q_{EW}^2}{s\xi^2_C} - I_\epsilon(k_m,k_n) \right] \, ,
\end{equation}
where the sum runs over all the massive--massive particle pairs. $I_0$ and $I_\epsilon$ are taken from Eq. (A.41) 
and Eq.~(A.50) of Ref.~\cite{Alioli:2010xd}, respectively.

For the real emission part, the subtraction formulae are the same as in QCD with the obvious substitutions 
$\alpha_s \rightarrow \aem$, $C_F \rightarrow q^2/q_iq_j$ (see Eq. (2.3) of Ref.~\cite{Dittmaier:1999mb}). The collinear remnants are the same 
as in Eq. (2.102) of Ref.~\cite{Frixione:2007vw} with the only substitutions necessary for the EW real part. 
Note that, since the collinear remnants contain a finite part after cancellation of all the singularities present in the NLO calculation 
and PDFs, this finite part is taken under control correctly only when using a PDF set accounting for 
both QCD and QED radiation off the quarks, like {\it e.g.} MRST2004QED~\cite{Martin:2004dh}, in order to be 
consistent with the full NLO calculation. However, it is known from various studies that the 
QED contribution to the PDF evolution is actually very small for the typical $x$ values contributing to the
DY process at collider energies. As a consequence, PDF sets describing QCD radiation only 
can be safely used in association with complete QCD/EW collinear remnants, as done in 
the numerical results shown in Section \ref{sec:results}.\footnote{Note also that, because we do not
consider in our study PDF sets describing the presence of photons inside the proton, we do not include
correspondingly the contribution of the so--called photon--induced processes among the real EW diagrams. 
However, they could be simply added in the calculation, provided an appropriate PDF set is used.}

It is worth noting that the implemented subtraction procedure is not strictly related to the 
CC DY process, but, in principle, 
it is generally valid for any process involving the interaction of charged
particles.

\section{Further issues}
\label{sec:implementation}
In the present Section, we briefly describe further modifications of
the \POWHEGBOX{} for an efficient and complete inclusion of EW
radiation contributions.
\subsection{Radiation phase space for FSR from massive partons}
\label{sec:fsrs}
In principle, we could have treated
the leptons as massless, using the same final state radiation
setup that is used for final state massless partons in the 
QCD version of the \POWHEGBOX{}. As mentioned earlier,
we have instead preferred to keep as finite the lepton mass in the 
treatment of final state
radiation from the leptons. Therefore, we had to introduce a new
final state mapping for FSR, to be used for 
partons with small mass. This mapping augments the possibilities
available in the \POWHEGBOX{}, described in detail in 
Section 5.1 (for ISR) and Section 5.2 (for FSR) of Ref.~\cite{Frixione:2007vw},
and is detailed in Appendix~\ref{sec:fsr} of the present work.
There were a few reasons to go along this direction. One reason
is that the EW virtual corrections, as previously described, are available with a finite mass of the
leptons. A second reason has to do with the fact
that the mass of the leptons is the true physical cutoff, and thus by
including the mass one has in fact a better result, that may
also be used safely for not so light leptons.
\subsection{Generation of the hardest radiation}
\label{sec:interface}
As already pointed out in Section~\ref{sec:powheg}, we may have up to three singular regions
associated with the real graphs sharing the same underlying
Born configuration. For example, the underlying Born flavour
$d\bar{u}\to e \bar\nu$ has three $\frindsing$ associated
with it, one for the $d\bar{u}\to e \bar\nu g$ real process,
with an ISR region, one for $d\bar{u}\to e \bar\nu \gamma$
in the ISR region, which is singular only when the transverse
momentum of the photon is small, and one for $d\bar{u}\to e \bar\nu \gamma$,
with a FSR mass singularity (in the mass of the lepton) when
the photon and lepton momenta become parallel.
The \POWHEG{} Sudakov form factor for radiation of Eq.~(\ref{eq:suddef}) is 
therefore equal to the product of the Sudakov form factors for each
of these regions.
The \POWHEGBOX{} handles it by generating
a radiation transverse momentum with each of these Sudakov form
factors, and then choosing the one with the largest transverse
momentum ({\it i.e.} using the so--called highest bid procedure).
In all cases, a lower radiation $\pt$--cutoff is needed, in
order for the generation to terminate in finite time.
This is set equal to a typical hadronic scale for gluon
or photon radiation from quarks, while it is taken as the
mass of the lepton for photon radiation off the leptons.

\section{Numerical results}
\label{sec:results}

To validate all the new theoretical and computational ingredients of the
\POWHEGBOX{} described above, we performed 
a number of detailed numerical simulations of $W$ hadroprodution at the LHC energies. 
More precisely
\begin{enumerate}

\item we compared the results of the \POWHEGBOX{} with the predictions of the \texttt{HORACE} generator, 
in the presence of NLO EW corrections only. Note that \texttt{HORACE} implements completely independent 
EW form factors and real photon ME, computed in the mass regularization scheme. It can be 
considered as a benchmark, having been validated against other codes and hadron collider data.

\item We provided full results of the \POWHEGBOX{} for NLO strong and EW corrections interfaced to MC showers, 
using \texttt{PYTHIA} and \texttt{PHOTOS} for QCD and QED showers, respectively. We compared these 
predictions for the combined QCD$\otimes$EW corrections with those quoted in 
Ref.~\cite{Balossini:2009sa}. Note that the results of Ref.~\cite{Balossini:2009sa} were obtained 
according to a different methodology, by combining the events generated
with \texttt{HORACE} (and showered with the \texttt{HERWIG} Parton Shower) with the 
QCD predictions of \texttt{MC@NLO}.

\end{enumerate}

We considered the process $p p \to W^+ \to \mu^+ \nu_{\mu} + (X)$, at the c.m. energy $\sqrt{s} = 7$~TeV 
and using the CTEQ6L PDF set with factorization/renormalization scale $\mu = M_W$ for the
comparisons with the \texttt{HORACE} code ($\mu = M_W$ being the default choice of \texttt{HORACE}) and
$\mu = M_{l\nu}$ (the lepton--neutrino invariant mass) for the combination of EW and QCD corrections.
We applied the 
following acceptance cuts:
\begin{eqnarray}
p_\perp^\mu, \, \rlap{\slash}{\! E_T}  \geq 25 \, {\rm GeV} \, , \qquad \qquad |\eta_\mu| \leq 2.5 \, ,
\end{eqnarray}
and considered ``bare" ({\it i.e.} without photon recombination) event selection conditions.

The results have been obtained in the $G_\mu$ scheme, using the following input parameters
\begin{center}
\begin{tabular}{lll}
$G_{\mu} = 1.16637~10^{-5}$ GeV$^{-2}$ & 
$M_W = 80.425$~GeV&
$M_Z = 91.1876$~GeV \\
$\Gamma_W = 2.093$~GeV & 
$\sin^2\theta_W = 1 - M_W^2/M_Z^2$&
$M_{\rm Higgs} = 115$~GeV\\
$m_e=510.99892$~KeV &
$m_{\mu}=105.658369$~MeV &
$m_{\tau}=1.77699$~GeV \\
$m_u = 66$~MeV &
$m_c = 1.2$~GeV &
$m_t = 178$~GeV \\
$m_d = 66$~MeV &
$m_s = 150$~MeV &
$m_b = 4.4$~GeV \\
$V_{ud}=\sqrt{1 - V_{cd}^2}$ &
$V_{us}=0.222$ &
$V_{ub}=0$ \\
$V_{cd}=0.222$ &
$V_{cs}= 0.975$ &
$V_{cb}=0$ \\
$V_{td}=0$ &
$V_{ts}=0$ &
$V_{tb}=1$ \\
\end{tabular}
\end{center}
For the coupling of external photons to charged particles needed for the evaluation of photonic corrections we use 
$\aem = \aem (0) =1/137.03599911$, and $\alpha_s (M_Z) = 0.118$ with NLO evolution
for the results about the QCD$\otimes$EW combination.

\begin{figure}[hbtp]
\begin{center}
\includegraphics[height=6.5cm]{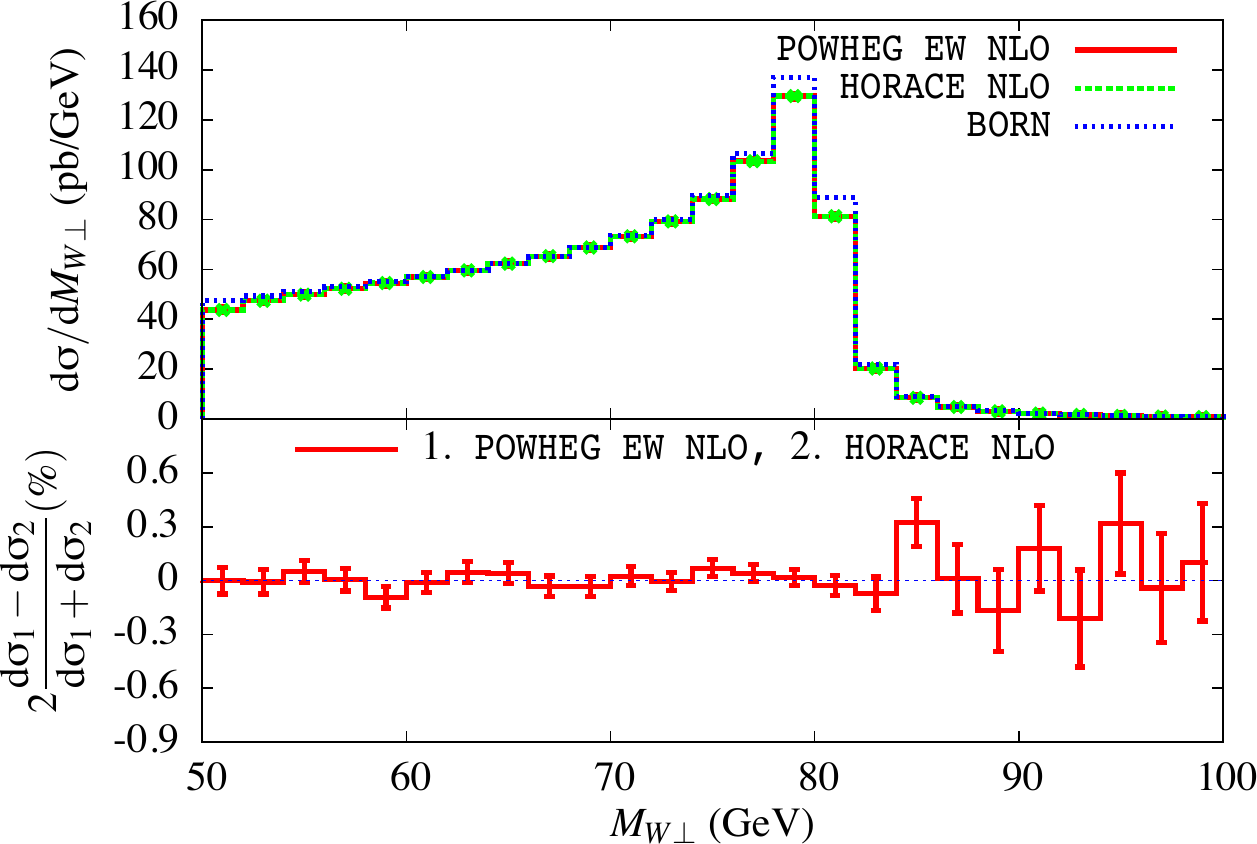}
\includegraphics[height=6.5cm]{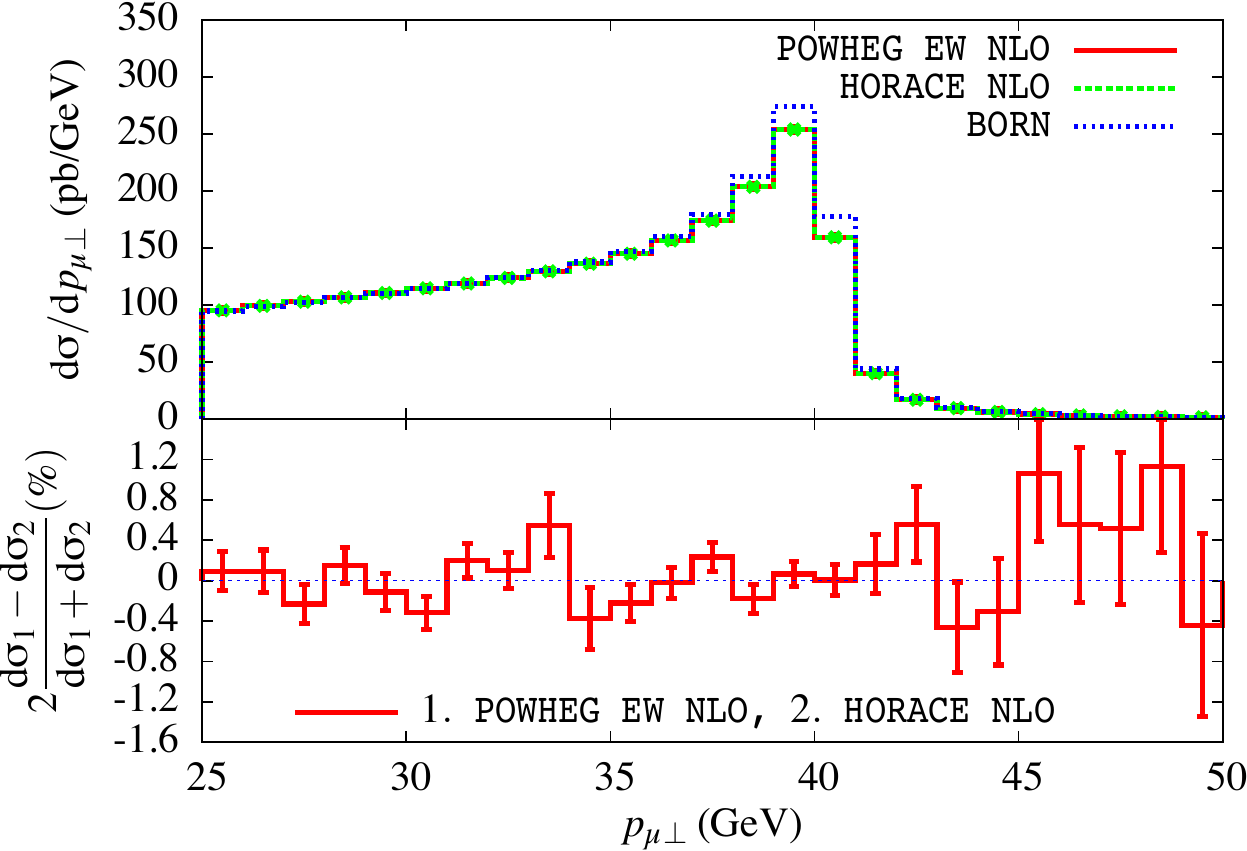}
\caption{Upper panels: the $W$ transverse mass (upper plot) and muon transverse momentum (lower plot) 
distributions according to the 
\POWHEGBOX{} and \texttt{HORACE} codes in the presence of 
pure NLO EW corrections, in the peak region. Lower panels: relative deviations, in per cent, between the 
predictions of the two generators.}
\label{ewpeak}
\end{center}
\end{figure}

\subsection{Electroweak comparisons with \texttt{HORACE}}

\begin{figure}[h]
\begin{center}
\includegraphics[height=6.5cm]{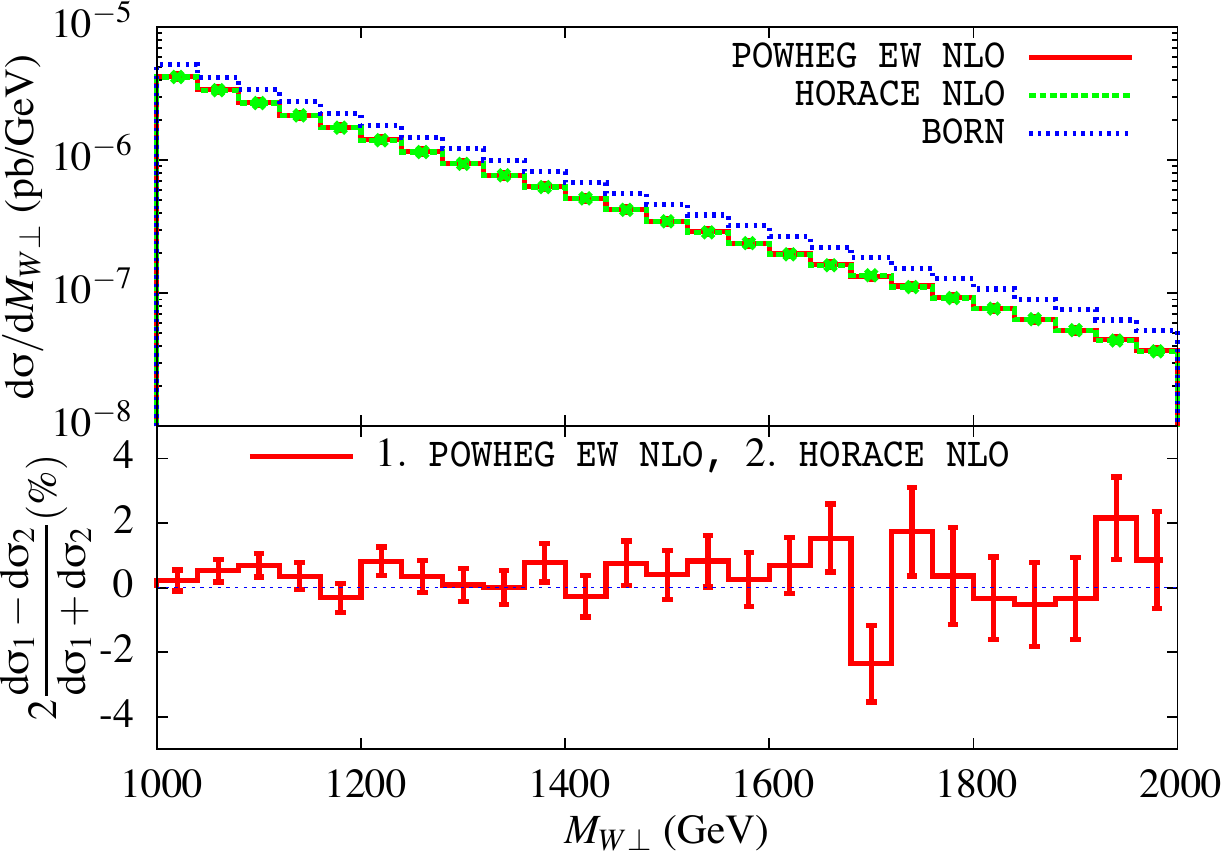}\\
\includegraphics[height=6.5cm]{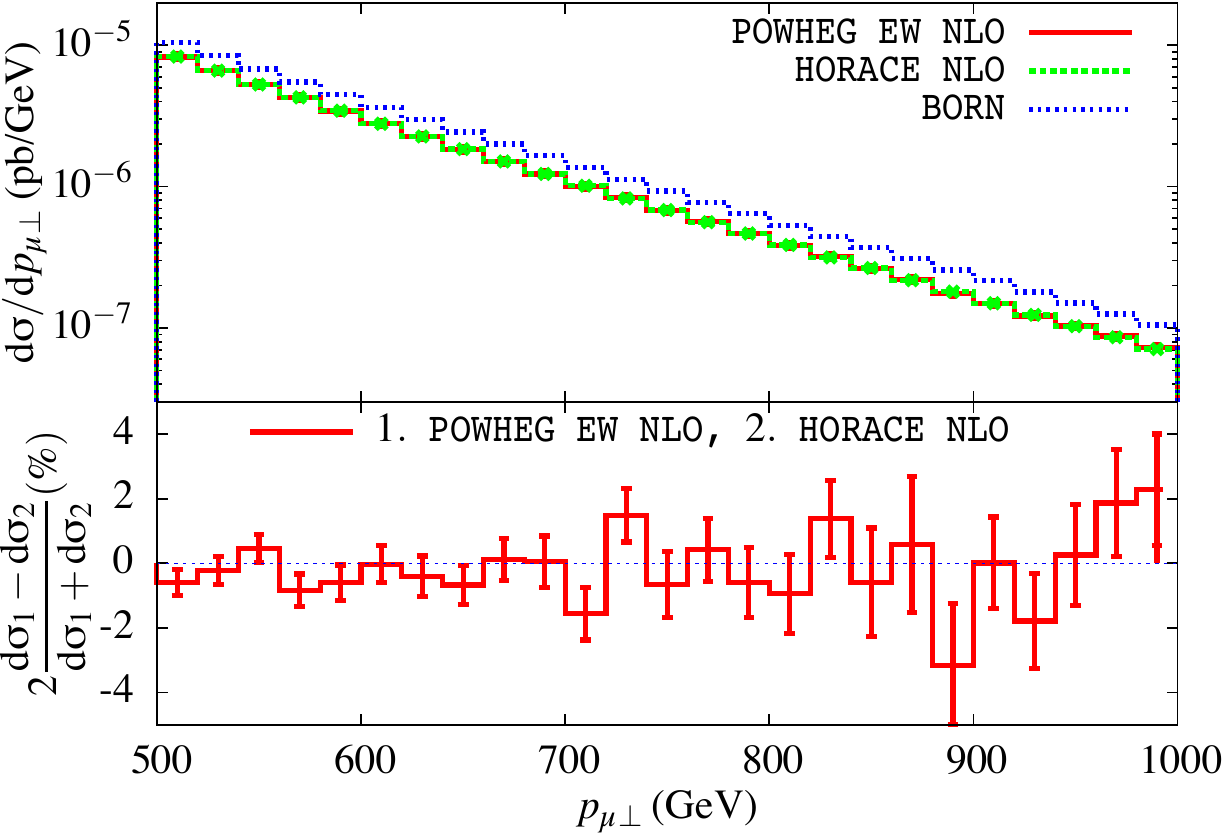}
\caption{The same as Fig. \ref{ewpeak} in the very high tail of the $M_\perp^W$ and lepton $p_\perp$ distributions.}
\label{ewfarpeak}
\end{center}
\end{figure}

The comparisons between the \POWHEGBOX{} and \texttt{HORACE} at the level of pure NLO EW corrections are shown in Fig.~\ref{ewpeak} and 
Fig.~\ref{ewfarpeak}, for the energy around the Jacobian peak and far from it, respectively. In the upper panels, the 
absolute predictions of the 
two codes for the $W$ transverse mass and lepton transverse momentum distributions are shown, together with the reference Born results.
The NLO EW predictions of the  \POWHEGBOX{}  have been obtained by setting in the code $\alpha_s$ at a fictitious, infinitesimal 
numerical value.
The typical some percent reduction of the peak cross section due to EW corrections is clearly seen for both the codes 
in Fig.~\ref{ewpeak}, while the significant effect due to the EW Sudakov logarithms can be appreciated in 
Fig.~\ref{ewfarpeak}. The lower panels shows the relative difference in per cent between the two generators. The error bars correspond to 1$\sigma$ MC 
errors. It can be seen that the results of the programs agree within the statistical fluctuations, both in the peak region and well above it.

This agreement represents a non--trivial check of the correctness of all the improvements and new features realized 
in the \POWHEGBOX{}, from the calculation of the EW virtual contributions to the improved subtraction procedure and the new FSR mapping 
of collinear photon singularities. 
Indeed, it is known that around the $W$ resonance the NLO EW corrections are largely dominated by 
final state photon radiation, containing
logarithms of the type $\log(\hat{s}/m_l^2)$, which are therefore correctly accounted for by the newly implemented FSR mapping
and generalized subtraction scheme.
On the other hand, in the high energy tails of the distributions, the NLO EW corrections display large Sudakov logarithms 
due to the
exchange of EW massive gauge bosons in the loops and yielding corrections of tens of per cent. The 
agreement between the two codes shown in Fig.~\ref{ewfarpeak} demonstrates the correct treatment of the NLO EW virtual contributions 
in the \POWHEGBOX{}.

We performed further numerical tests not shown here, in particular on the size and shape of the NLO EW corrections to 
various observables for both the muon and electron final state, 
to register perfect technical agreement with \texttt{HORACE}.

\subsection{Combined electroweak and QCD corrections}

\begin{figure}[hbtp]
\begin{center}
\includegraphics[height=7.25cm]{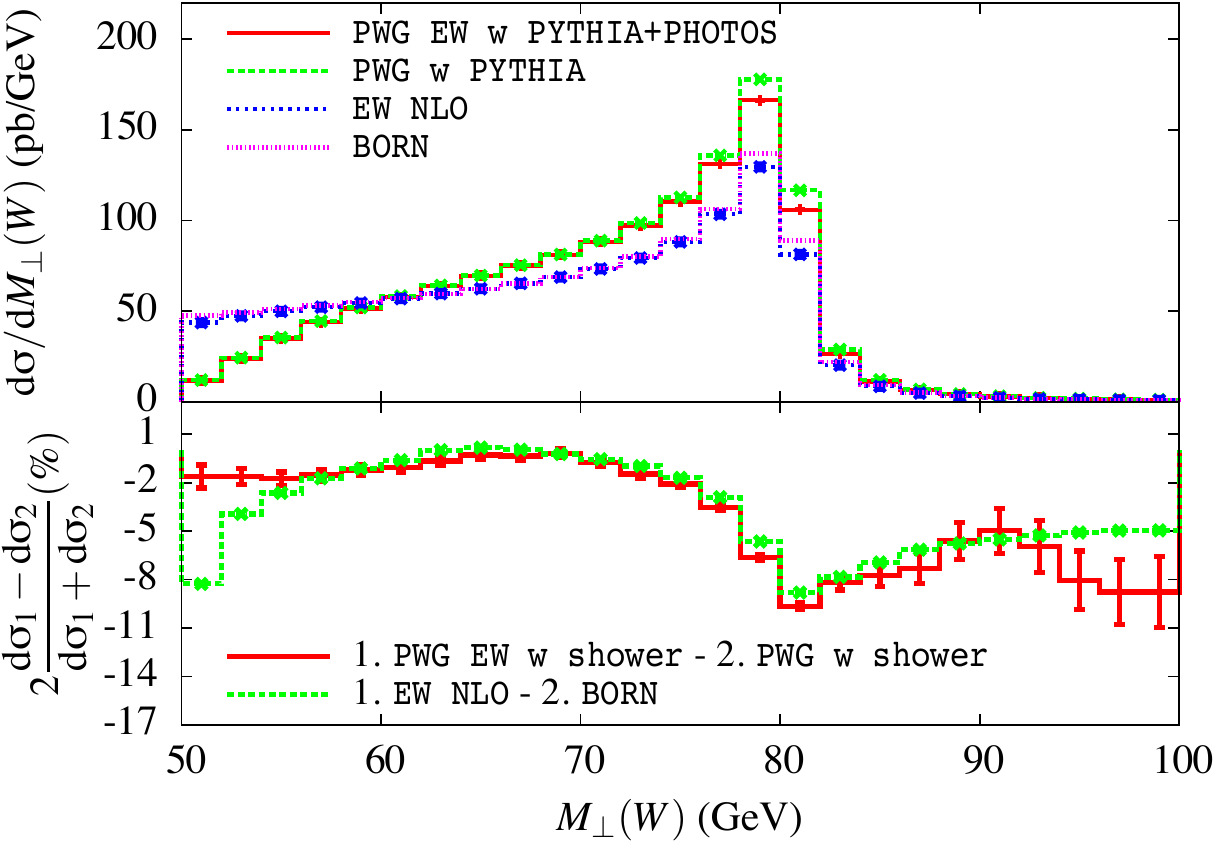}
\caption{Upper panel: the $W$ transverse mass distribution according to the full QCD$\otimes$EW predictions of the 
\POWHEGBOX{} (denoted as {\tt PWG EW w PYTHIA+PHOTOS}), 
the standard QCD \POWHEGBOX{} ({\tt PWG w PYTHIA}), the LO and the NLO EW approximations. 
Lower panel: relative difference, in per cent, between the full QCD$\otimes$EW predictions and 
the pure QCD ones (red, solid line), in comparison with the relative contribution due to pure NLO EW corrections (green, dotted line). 
The difference between the two lines is
a measure of the mixed QCD$\otimes$EW corrections.}
\label{mtwqcdew}
\end{center}
\end{figure}

\begin{figure}[hbtp]
\begin{center}
\includegraphics[height=7.25cm]{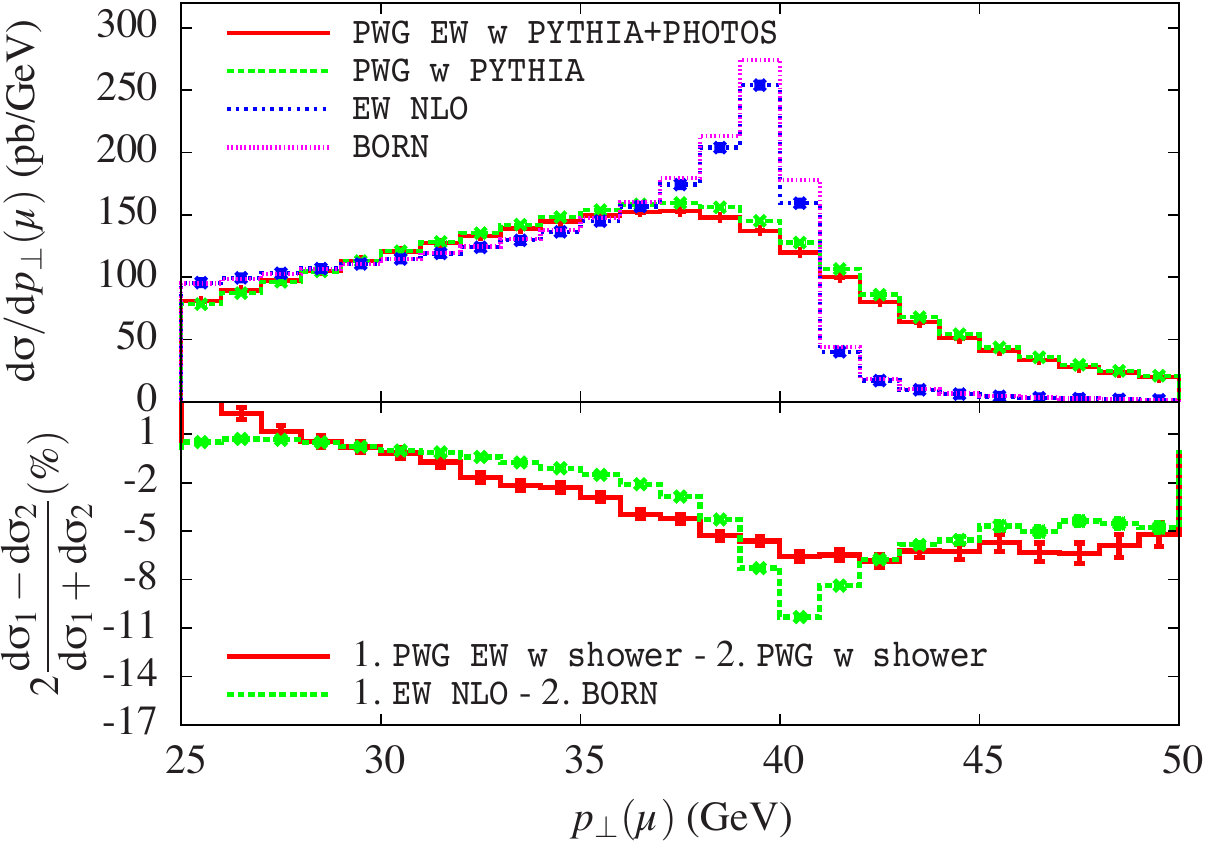}
\caption{The same as Fig.~\ref{mtwqcdew} for the lepton transverse momentum distribution.}
\label{ptlqcdew}
\end{center}
\end{figure}

\begin{figure}[hbtp]
\begin{center}
\includegraphics[height=7cm]{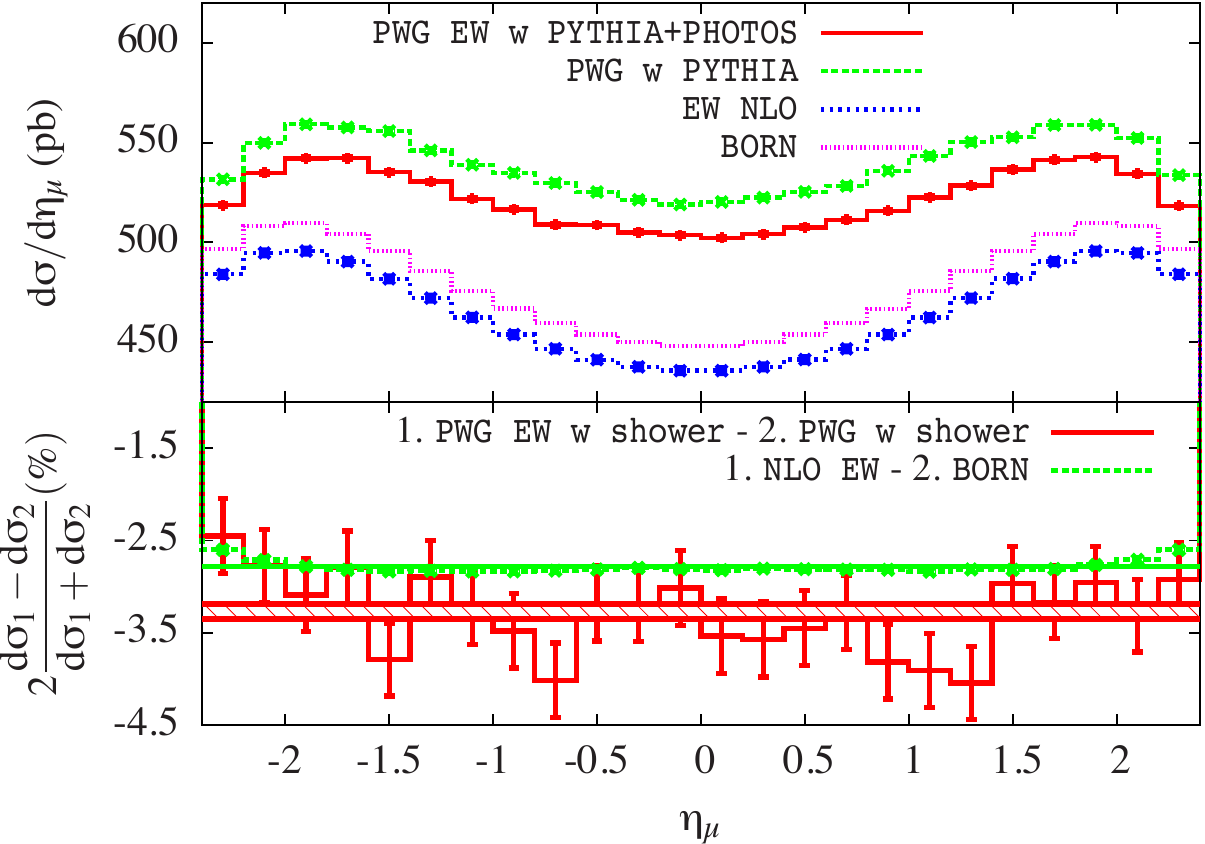}
\caption{The same as Fig. ~\ref{mtwqcdew} for the lepton pseudorapidity distribution.}
\label{rapidityqcdew}
\end{center}
\end{figure}

\begin{figure}[hbtp]
\begin{center}
\includegraphics[height=7cm]{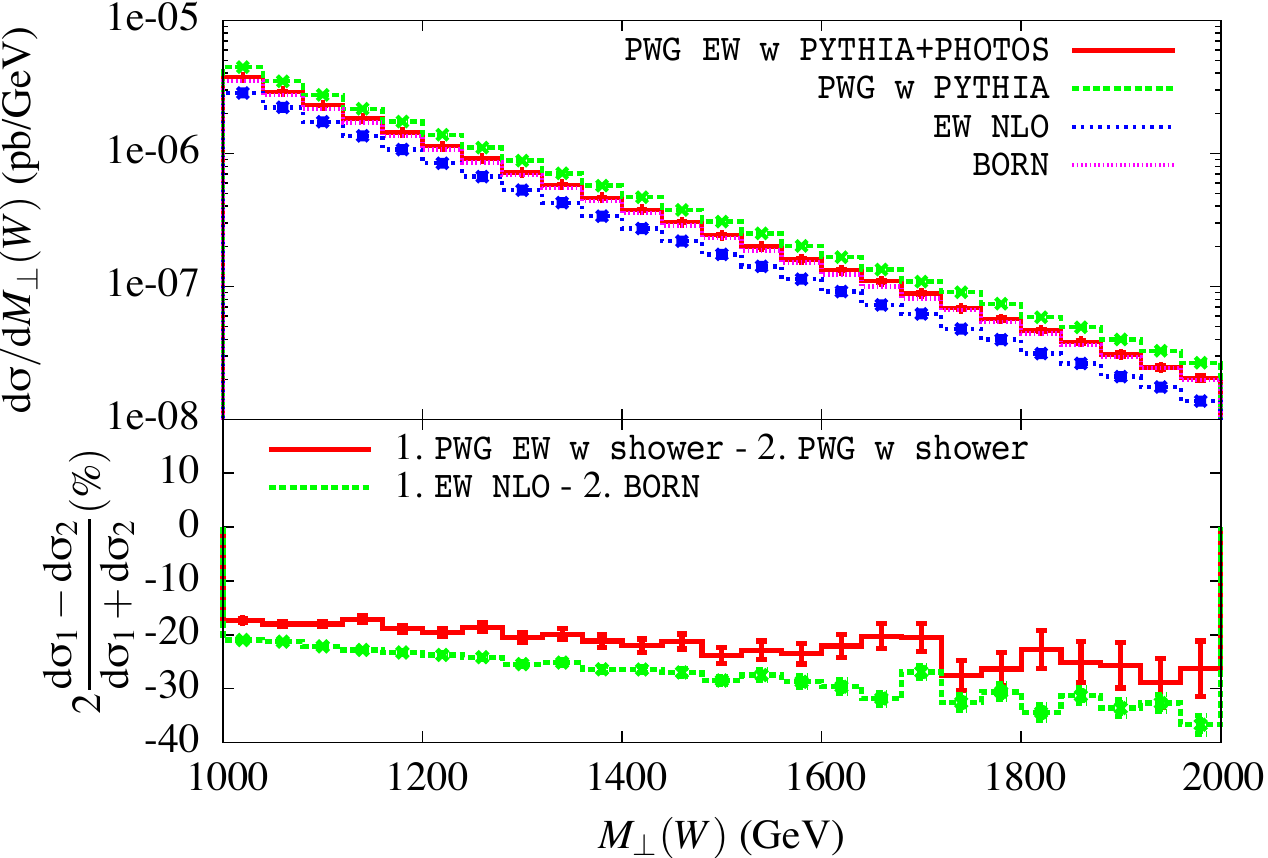}
\caption{The same as Fig.~\ref{mtwqcdew} for the transverse mass distribution far from the resonance region.}
\label{ewqcd-highmass}
\end{center}
\end{figure}

The full results of the \POWHEGBOX{} for the combined effects of QCD and EW radiation in the resonance region are shown in 
Fig.~\ref{mtwqcdew} for the 
$W$ transverse mass, in
Fig.~\ref{ptlqcdew} for the lepton $p_\perp$
and in Fig.~\ref{rapidityqcdew}  for the lepton rapidity.
In Fig.~\ref{ewqcd-highmass} we show
the transverse mass distribution above 1~TeV. The full predictions have been obtained by interfacing the NLO 
EW and strong corrections with 
QCD (\texttt{PYTHIA}) and QED (\texttt{PHOTOS}) showers.  For the sake of 
comparison, the pure QCD predictions of the standard 
\POWHEGBOX{} are also given, together with the 
pure NLO EW results. These absolute predictions for the various distributions are shown in the upper panels of each plot. 
The lower panels display the relative difference, in per cent, between the results of the new version of the \POWHEGBOX{} 
and the standard QCD release, as well as the relative effect due to pure NLO EW corrections. Therefore the comparison 
between the two lines in each lower panel provides a measure of the combination of QCD and EW corrections 
and, more precisely, of mixed EW$\otimes$QCD contributions at 
order $\aem^n \alpha_s^n, n \geq 1$ in perturbation theory.\footnote{The exact ${\cal O}(\aem\alpha_s)$ corrections to 
DY processes are presently unknown, albeit partial results 
are available in the literature~\cite{Kuhn:2007qc,Kuhn:2007cv,Hollik:2007sq,Denner:2009gj,Kilgore:2011pa} and further work is in progress along this direction.}

From the upper panels, one can clearly see that NLO QCD corrections in association with QCD shower effects are strictly needed for a correct simulation of both the
normalization and shape of the distributions. Especially for the lepton $p_\perp$, the particularly pronounced smearing of the LO prediction and the rising 
of a heavy tail above the Jacobian peak, that are the well known effects due to QCD 
radiation, are clearly visible. However, also EW radiation, when combined with QCD effects, plays a role for precise calculations of the distributions, impacting both on the normalization and shape of the distributions themselves.

The latter observation can be better understood by looking 
in detail at the lower panels of  Figs.~\ref{mtwqcdew}--\ref{ewqcd-highmass}. 
Actually, the two lines show, as already emphasized, 
the relative contribution due to EW radiation in association with QCD effects and the relative contribution due to NLO EW contributions only. For the $W$ transverse mass around the peak (Fig.~\ref{mtwqcdew}), the substantial agreement between the two lines indicates that the relative impact of the EW 
corrections is not significantly altered by the combination with QCD radiation. 
Stated differently, both corrections are necessary because the mainly positive and dominant QCD effects are partially compensated by the negative EW contributions, 
but their interplay gives rise to a relative contribution on the same ground as that provided by a pure NLO EW calculation. This means that, not surprisingly, 
the impact due to mixed EW$\otimes$QCD corrections to the $W$ transverse mass is quite moderate, 
with the exception of the region $M_\perp^W$ around $M_W/2$, as a consequence of the 
rather mild dependence of such a 
distribution on QCD effects.

However, the same conclusions can not be drawn for the muon transverse momentum and rapidity distributions (Fig.~\ref{ptlqcdew} and Fig.~\ref{rapidityqcdew}) 
in the peak region and for the transverse mass in the high tail (Fig.~\ref{ewqcd-highmass}). For 
the lepton $p_\perp$, the two lines in the lower panel show a 
systematic, statistically significant difference, particularly around the Jacobian peak. In this region the particularly large QCD corrections 
to the lepton $p_\perp$ conspire with the per cent level EW 
contributions to give rise to non--negligible mixed corrections at a few per cent level. Also for the lepton rapidity, as shown in 
Fig.~\ref{rapidityqcdew}, the negative EW corrections partially reduce the positive effect of QCD radiation, yielding combined 
QCD$\otimes$EW contributions of about one per cent. A similar reasoning applies to the transverse mass above 1~TeV shown in Fig.~\ref{ewqcd-highmass}, where the quite large EW corrections, enhanced by Sudakov logarithms, in association with QCD radiation effects produce mixed contributions changing the normalization of some per cent.

We compared the results discussed in this Section, as well as our predictions for other observables like the $W$ rapidity, with those 
derived in Ref.~\cite{Balossini:2009sa} for the 
combination of EW and strong corrections to $W$ production. In spite of some different details used in the simulations of 
Ref.~\cite{Balossini:2009sa} (different PDF set, event selection cuts 
and input parameters) we observed 
a satisfactory agreement
for the size and shape of the effects due to the QCD$\otimes$EW combination. However, whereas in Ref.~\cite{Balossini:2009sa} the study required, as still currently done in the simulations performed by the experimental collaborations at the Tevatron and the LHC, a rather time--consuming use of a tandem of generators 
(\texttt{HORACE}+\texttt{HERWIG} and \texttt{MC@NLO}) and the 
merging of their output, the results here given were 
directly obtained by means of a single computational framework.

\section{Conclusions}
\label{sec:concl}

We realized the first fully consistent implementation of EW radiation effects in the \POWHEGBOX{}. We considered the process of single 
$W$ hadroproduction as a case study and 
described how NLO EW corrections have been combined with the already available NLO QCD calculation. We also interfaced NLO EW and QCD corrections with 
QCD and QED MC showers. The resulting tool allows to study comprehensively the interplay of QCD and EW effects to $W$ production at hadron colliders 
according to a unified framework, thus facilitating the work of MC simulations in data analysis.

We presented several successful cross--checks of our predictions against benchmark results to emphasize the reliability and accuracy of the new tool.

The approach followed for the implementation of the EW effects mimicked the general, process independent structure of the 
\POWHEGBOX{} framework. It therefore 
paves the way for an almost straightforward inclusion in the \POWHEGBOX{} of EW radiation effects in further processes, like single $Z$ hadroprodution and other processes of
phenomenological interest. 

These perspectives are left to future works.

\vskip 12pt\noindent
{\bf \large Note added in proof}
\vskip 8pt\noindent
During the completion of this work, an independent combination of NLO
QCD and EW corrections to $W$ hadroproduction in the POWHEG BOX framework
appeared in Ref.~\cite{Bernaciak:2012hj}. This study differs from our approach in several
aspects, the most important one being the fact that multiple photon
radiation in not dealt with in Ref.~\cite{Bernaciak:2012hj}. It would be interesting to compare
the results of the two implementations in the future.

\acknowledgments
We are grateful to Carlo Oleari, Giacomo Polesello and Torbjorn Sjostrand for useful discussions and correspondence. \\
This work was supported in part 
by the Research Executive Agency (REA) of the European Union under the Grant Agreement number PITN-GA-2010-264564 (LHCPhenoNet).

\appendix

\newcommand{\tmtextrm}[1]{{\rmfamily{#1}}}

\section{Final state radiation mapping for massive partons} \label{sec:fsr}

In this Appendix we document the construction of the final state radiation
machinery for the radiation of a massless parton off a massive one. We refer
throughout this Appendix to the notation of Ref.~{\cite{Frixione:2007vw}},
where the $\Phi_n$ refers to the Born phase space, and $\Phi_{n + 1}$ refers
to the real radiation phase space. The $n$ and $n + 1$ partons are the
radiating and radiated partons, respectively. The real momenta are written $k_1
\ldots k_{n + 1}$ and the underlying Born momenta are $\bar{k}_1 \ldots
\bar{k}_n$. We will deal here with the case when the $n^{\tmop{th}}$ parton is
massive.

The main ingredient for this construction are the following:
\begin{itemize}
  \item the construction of a factorized phase space $d \Phi_{n + 1} = d
  \Phi_n \hspace{0.25em} d \Phi_{\tmop{rad}}$, where the full $(n + 1)$
  particle phase space is expressed in terms of a $n$ particle phase space and
  a (three dimensional) radiation phase space. This factorization must be
  unique in both direction, {\it i.e.} given a $\Phi_{n + 1}$ point we should get a
  unique $(\Phi_n, \Phi_{\tmop{rad}})$ pair and viceversa. Once the factorized
  phase space is known, we can compute the $\bar{B}$ function in
  \tmtexttt{POWHEG}.
  
  \item A definition of a hardness scale for radiation $K_{\perp}^2$ should be
  given, that coincides with the usual definition in the massless limit. This
  scale will appear in the Sudakov form factor inside a theta function $\theta
  (K_{\perp}^2 - p^2)$, where $p$ is the argument of the Sudakov form factor.
  It should be chosen in such a way that the integral
  $d \Phi_{\tmop{rad}}\theta  (K_{\perp}^2 - p^2)$
  should be easy to do.
  
  \item A form for an upper bounding function of $R / B$ must be given for the
  generation of radiation, such that the associated Sudakov form factor can be
  computed analytically. In order for this to be possible, this upper bounding
  function must be integrable over the radiation phase space with the theta
  function $\theta (K_{\perp}^2 - p^2)$. Since we are seeking for an upper
  bound to $R / B$, we may extend the phase space $\Phi_{\tmop{rad}}$ beyond
  its kinematic limits if that makes the integration easier, assuming later
  that $R / B$ vanishes when the radiation variables are outside 
  the kinematic bounds.
  
  \item The integral of the upper bounding function with $\theta
  (K_{\perp}^2 - p^2)$ over the (possibly extended) radiation phase
  space should be performed analytically. Furthermore, once a value of $p$ is
  generated using the Sudakov form factor constructed with the upper bound of
  $R / B$, the generation of  $\Phi_{\tmop{rad}}$ with the constraint
  $K_\perp=p$ should also be set up. Once the full $\Phi_{\tmop{rad}}$ point is
  generated, one can use the veto technique to implement the real Sudakov form
  factor.
\end{itemize}

\subsection{Factorized phase space}

As in the massless case, the underlying Born configuration will be constructed
as follows. First of all, we work in the c.m. frame of the $k_1
\ldots k_{n + 1}$ system. We define
\begin{equation}
  q = \sum_{i = 1}^{n + 1} k_i
\end{equation}
and we will use $q$ also to denote $\sqrt{q^2}$, when this does not generate
confusion.

We define the underlying Born $\bar{k}_n$ to be (spatially) parallel to $k_n$.
The underlying Born recoil system $\bar{k}_1 \ldots \bar{k}_{n - 1}$ is
obtained by boosting the $k_1 \ldots k_{n - 1}$ recoil system in a direction
parallel to $k_n$, in such a way that its three--momentum balances the
$\bar{k}_n$ three--momentum. The modulus of the $\bar{k}_n$ three--momentum is
chosen in such a way that the total energy of the $\bar{k}_1 \ldots \bar{k}_n$
system equals the total energy of the $k_1 \ldots k_{n + 1}$ system. Defining
$k_{\text{\tmtextrm{rec}}} = \sum_{i = 1}^{n - 1} k_i $ and
$\bar{k}_{\text{\tmtextrm{rec}}} = \sum_{i = 1}^{n - 1} \bar{k}_i $, we get
immediately
\begin{equation}
  \bar{k}_n^0 = \frac{q^2 + m^2 - M_{\tmop{rec}}^2}{2 q},
  \, \, \bar{k}^0_{\tmop{rec}} = \frac{q^2 - m^2 + M_{\tmop{rec}}^2}{2 q},
  \, \, M_{\tmop{rec}}^2 \equiv k_{\text{\tmtextrm{rec}}}^2 =
  \bar{k}_{\text{\tmtextrm{rec}}}^2 .
\end{equation}
The construction of the radiation phase space is slightly more involved, and
it proceeds as follows.

We begin by writing the $n + 1$ body phase space in factorized form of a three--body 
phase space involving the $n^{\text{\tmtextrm{th}}}$ and $(n +
1)^{\text{\tmtextrm{th}}}$ partons, and the recoil system momentum
$k_{\text{\tmtextrm{rec}}}$, times the phase space $d \Phi_{\tmop{rec}}$ of
the recoil system at fixed $k_{\text{\tmtextrm{rec}}}$
\begin{eqnarray}
  d \Phi_{n+1}&=& d \Phi_3 \,d\Phi_{\rm rec}\,, \label{eq:np1fact} \\
  d \Phi_3 &=& \frac{dM_{\tmop{rec}}^2}{2 \pi} \frac{d^3 k_{n + 1}}{2 k_{n
  + 1}^0 (2 \pi)^3} \frac{d^3 k_n}{2 k_n^0 (2 \pi)^3} \frac{d^3
  k_{\tmop{rec}}}{2 k_{\tmop{rec}}^0 (2 \pi)^3} (2 \pi)^4 \delta^4 (q - k_{n +
  1} - k_n - k_{\tmop{rec}}) .
\end{eqnarray}
The three--body phase space part can be written in term of the Dalitz variables:
\begin{eqnarray}
 d \Phi_3 &=& 
 \frac{d^3 k_{n + 1}}{2 k_{n + 1}^0 (2 \pi)^3} \frac{d^3 k_n}{2 k_n^0 (2
   \pi)^3} 2 \pi \delta ((q - k_{n + 1} - k_n)^2 - M_{\tmop{rec}}^2)\nonumber \\
 &=& \frac{d \Omega_{n + 1}}{4 (2 \pi)^6} k_{n + 1} dk_{n + 1}^0 k_n dk_n^0 d
   \cos\theta d \phi 2 \pi \delta ((q - k_{n + 1} - k_n)^2 -
   M_{\tmop{rec}}^2) .
\end{eqnarray}
Now
\[ (q - k_{n + 1} - k_n)^2 - M_{\tmop{rec}}^2 = q^2 + m_n^2 - M_{\tmop{rec}}^2
   - 2 q^0 (k_{n + 1}^0 + k_n^0) + 2 k_{n + 1}^0 k_n^0 + 2 \cos \theta k_{n +
   1} k_n, \]
so that the $\delta$ function can be integrated in $d\cos\theta$,
yielding
\begin{equation}
  d \Phi_3 = \frac{d \Omega_{n + 1}}{8 (2 \pi)^5} dk_{n + 1}^0 dk_n^0 d \phi \, .
\end{equation}
The orientation can be taken relative to any of the three bodies, so we can
as well write
\begin{equation}
  d \Phi_3 = \frac{d \Omega}{8 (2 \pi)^5} dk_{n + 1}^0 dk_n^0 d \phi \, ,
\end{equation}
where $\Omega$ is the direction of $k_{\tmop{rec}}$, and $\phi$ is the azimuth
of $k_n$ or $k_{n + 1}$ relative to $k_{\tmop{rec}}$.

The underlying Born phase space can be factorized into a two--body phase space
times the phase space of the system recoiling against the emitting parton
\begin{equation}
  d \bar{\Phi}_n = \frac{dM_{\tmop{rec}}^2}{2 \pi} \frac{d^3 \bar{k}_n}{2
  \bar{k}_n^0 (2 \pi)^3} \frac{d^3 \bar{k}_{\tmop{rec}}}{2
  \bar{k}_{\tmop{rec}}^0 (2 \pi)^3} (2 \pi)^4 \delta^4 (q - \bar{k}_n -
  k_{\tmop{rec}}) d \bar{\Phi}_{\tmop{rec}} \, . \label{eq:nfact}
\end{equation}
We have
\begin{equation}
  \frac{d^3 \bar{k}_n}{2 \bar{k}_n^0 (2 \pi)^3} \frac{d^3 k_{\tmop{rec}}}{2
  k_{\tmop{rec}}^0 (2 \pi)^3} (2 \pi)^4 \delta^4 (q - \bar{k}_n -
  k_{\tmop{rec}}) = \frac{d^3 \bar{k}_n}{2 \bar{k}_n^0 (2 \pi)^3} 2 \pi \delta
  \left( (q - \bar{k}_n)^2 - M_{\tmop{rec}}^2 \right) = \frac{d \Omega}{32
  \pi^2} \frac{2 \bar{k}_n}{q} .
\end{equation}
We wish to express the phase space in terms of the underlying Born phase
space and the radiation variables, that we take equal to $k^0_{n + 1}$,
$k^0_n$ and $\phi$. In other words, we must identify Eq.~(\ref{eq:np1fact})
with Eq.~(\ref{eq:nfact}) times $dk^0_{n+1}dk^0 d\phi$ times
a jacobian $J$ that we should infer. This yields
\begin{eqnarray}
  \frac{d \Omega}{8 (2 \pi)^5} dk_{n + 1}^0 dk_n^0 d \phi
  \frac{dM_{\tmop{rec}}^2}{2 \pi} d \Phi_{\tmop{rec}} & = & Jdk_{n + 1}^0
  dk_n^0 d \phi d \bar{\Phi}_n \nonumber\\
  & = & Jdk_{n + 1}^0 dk_n^0 d \phi \frac{d \Omega}{32 \pi^2} \frac{2
  \bar{k}_n}{q} \frac{dM_{\tmop{rec}}^2}{2 \pi} d \bar{\Phi}_{\tmop{rec}} \, .
\end{eqnarray}
Since $\overline{\Phi_{}}_{\tmop{rec}}$ is obtained by boosting
$\Phi_{\tmop{rec}}$, the corresponding phase space elements cancel on both
sides, so that, canceling other common factors, we get
\begin{equation}
  J = \frac{1}{(2 \pi)^3} \frac{q}{2 \bar{k}_n},
\end{equation}
and
\begin{equation}
  d \Phi_{n + 1} = Jdk_{n + 1}^0 dk_n^0 d \phi d \bar{\Phi}_n .
\end{equation}
We thus define
\begin{equation}
  d \Phi_{\tmop{rad}} = Jdk_{n + 1}^0 dk_n^0 d \phi \, ,
\end{equation}
which solves the problem.

\begin{figure}[h]
\begin{center}
  \includegraphics[width=0.625\textwidth]{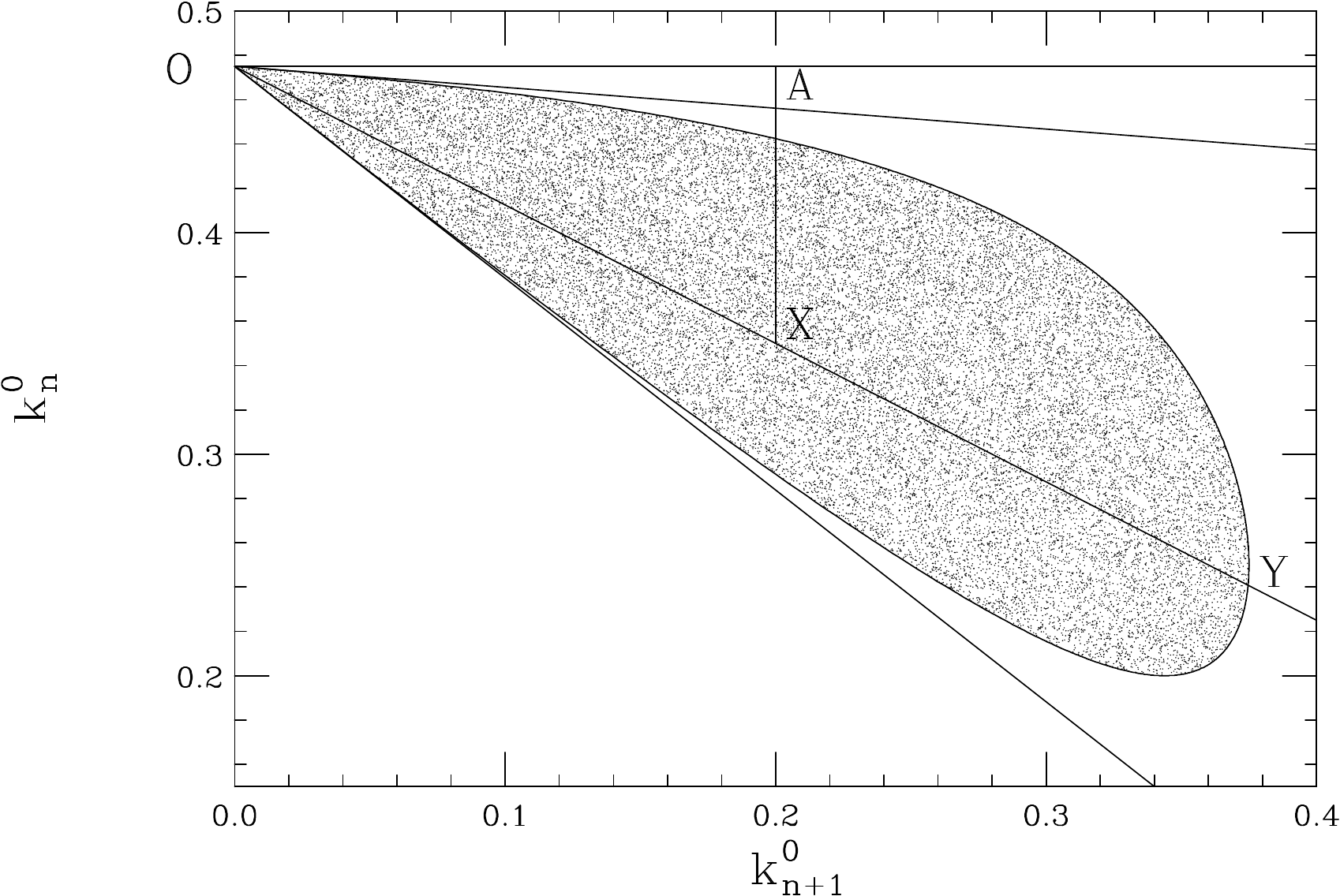}
  \caption{Dalitz region for $M_{\tmop{rec}} = 0.3$, $m = 0.2$ .}
   \label{dalitz}
\end{center}
\end{figure}

It is more convenient to introduce variables that have a closer
parallel to the FKS viariables. It turns out that $k^0_{n + 1}$, $k_n^0$ and
$k_{\tmop{rec}}^0$ live in a convex Dalitz domain shown in 
Fig.~\ref{dalitz}.
The boundary of that domain is defined by the configuration regions where the
three vectors $\vec{k}_{n + 1}$, $\vec{k}_n$ and $\vec{k}_{\tmop{rec}}$ lie on
the same line. The point at $k_{n + 1} = 0$ certainly belongs to that domain.
Notice that when $k_{n + 1} = 0$ we have $k^0_n = \bar{k}_n^0$ and
$k^{0_{}}_{\tmop{rec}} = \bar{k}^0_{\tmop{rec}}$. Since the Dalitz domain is
convex, it should be possible to parametrize it as a function of two
parameters, $k^0_{n + 1}$ itself and $z$, with
\begin{equation}
  k^0_n = \bar{k}^0_n - zk^0_{n + 1} \, . \label{eq:kfromy}
\end{equation}
This corresponds, in Fig. \ref{dalitz}, to parametrize the point $X$ by giving
the \ $k^0_{n + 1}$ value, and the tangent $y$ of the angle $\widehat{XOA}$.
Thus, the point $k_n^0 = \bar{k}^0_n$, $k_{n + 1} = 0$ belongs to the Dalitz
region for all values of $z$. For a given value of $z$, there is then a
maximum value of $k_{n + 1}$, such that the point is on the boundary of the
Dalitz region (the point $Y$ in Fig. \ref{dalitz}). It is characterized by the
condition
\begin{equation}
  | \vec{k}_{n + 1} | \pm | \vec{k}_n | \pm | \vec{k}_{\tmop{rec}} | = 0 \, ,
  \label{eq:triangleeq}
\end{equation}
that has to hold for at least one sign combination. Eq. (\ref{eq:triangleeq})
corresponds to the boundary of a triangular inequality. It is solved by
squaring
\begin{equation}
  (| \vec{k}_{n + 1} | \pm | \vec{k}_n |)^2 = k_{n + 1}^2 + \vec{k}_n^2 \pm 2
  k_{n + 1} | \vec{k}_n | = \vec{k}_{\tmop{rec}}^2 \, ,
\end{equation}
from which it follows again
\begin{equation}
  \left( k^2_{n + 1} + \vec{k}^2_n - \vec{k}^2_{\tmop{rec}} \right)^2 = 4 k_{n
  + 1}^2 \vec{k}_n^2 \, .
\end{equation}
We can now use $\vec{k}^2_n = {k^0_n}^2 - m^2$, $\vec{k}_{\tmop{rec}}^2 = (q -
k^0_n - k^0_{n + 1})^2 - M_{\tmop{rec}}^2$, and Eq. (\ref{eq:kfromy}), and we
obtain the equation for $k_{n + 1}$ of the form
\begin{equation}
  4 k_{n + 1}^2 \left( 2 k_{n + 1} qz (1 - z) + q^2 z^2 - 2 q
  \hat{k}^0_{\tmop{rec}} z + M_{\tmop{rec}}^2 \right) = 0 \, .
\end{equation}
This yields a double solution $k_{n + 1} = 0$, that we already knew, and
\begin{equation}
  k_{n + 1} = \frac{2 q \bar{k}^0_{\tmop{rec}} z - q^2 z^2 -
  M_{\tmop{rec}}^2}{2 qz (1 - z)} \, , \label{eq:knp1max}
\end{equation}
which is the sought maximum value of $k_{n + 1}$. The numerator on the r.h.s. 
of Eq. (\ref{eq:knp1max}) vanishes for
\begin{equation}
  z_{1 / 2} = \left( \bar{k}^0_{\tmop{rec}} \pm \sqrt{(
  \bar{k}^0_{\tmop{rec}})^2 - M_{\tmop{rec}}^2} \right) / q \, .
\end{equation}
These correspond to the maximum and minimum $z$ values allowed. The lines at
fixed $z = z_{1 / 2}$ are the lower/upper tangent to the Dalitz region from
the point $O$, shown in Fig.~\ref{dalitz}

We now define
\begin{equation}
  k_{n + 1} = \frac{\xi q}{2}, z = z_2 - (z_2 - z_1) (y + 1) / 2 \, ,
\end{equation}
where now $z$ and $y$ will play the role of the FKS variables. Notice that $y
= 1$ corresponds to $z = z_1$, which is the lower tangent in the plot. The
upper tangent corresponds to $k_n$ nearly constant, {\it i.e.} to $k_n$ recoiling
against $k_{n + 1}$ and $k_{\tmop{rec}}$, that are parallel. The lower tangent
corresponds instead to $k_{\tmop{rec}}$ recoiling against $k_{n + 1}, k_n$,
that are collinear. Thus, $y = 1$ corresponds to the mass singularity, which
is what we want.

Summarizing, the factorized phase space is given by
\begin{equation}
  d \Phi_{n + 1} = d \overline{\Phi_{}} d \Phi_{\tmop{rad}}, d
  \Phi_{\tmop{rad}} = \frac{1}{(2 \pi)^3} \hspace{0.25em} \frac{q^2}{4
  \bar{k}_n} k_{n + 1}^0 d \xi \hspace{0.25em} dz \hspace{0.25em} d \phi \, ,
\end{equation}
\begin{equation}
  k^0_{n + 1} = \frac{\xi q}{2}, k^0_n = \bar{k}^0_n - zk^0_{n + 1} \, ,
  \label{eq:zandcsi}
\end{equation}
where
\begin{equation}
  \bar{k}_n^0 \equiv \frac{q^2 + m^2 - M_{\tmop{rec}}^2}{2 q} \, .
\end{equation}
The physical region in $z$, $\phi$, $\xi$ is delimited as follows: $0 < \phi <
2 \pi$, \ $z_1 < z < z_2$, with
\begin{equation}
  z_{1 / 2} = \left( \bar{k}^0_{\tmop{rec}} \pm \sqrt{(
  \bar{k}^0_{\tmop{rec}})^2 - M_{\tmop{rec}}^2} \right) / q \, ,
  \bar{k}^0_{\tmop{rec}} \equiv \frac{q^2 - m^2 + M_{\tmop{rec}}^2}{2 q} \, .
  \label{eq:z12kbrec}
\end{equation}
Within \tmtexttt{POWHEG} $z$ is parametrized as
\begin{equation}
  z = z_2 - \frac{(1 + y)}{2} (z_2 - z_1) \, ,
\end{equation}
with $- 1 \leqslant y \leqslant 1$ playing the role of the usual FKS variable.
Furthermore, $k_{n + 1}$ satisfies the $z$--dependent bound
\begin{equation}
  0 \leqslant k_{n + 1} \leqslant \frac{2 q \bar{k}^0_{\tmop{rec}} z - q^2 z^2
  - M_{\tmop{rec}}^2}{2 qz (1 - z)} \, .
\end{equation}

\subsection{Definition of the hardness scale}

We need to find an appropriate definition for the hardness of the emission. In
the massless case, the transverse momentum is used, but this is not
appropriate now, because, in the soft limit, the maximum virtuality available
for the emitted gluon is no longer limited by the real transverse
momentum.\footnote{The maximum virtuality of the gluon is the scale to be used
for the strong coupling and thus, in this sense, it is the hardness of the
emission.} The singularity in the propagator of the massive quark has the
structure
\begin{equation}
  \frac{1}{(k + p)^2 - m^2} \, , \label{eq:massiveden}
\end{equation}
where we have called for ease of notation $k_{n + 1} \rightarrow k$, $k_n
\rightarrow p$. The hardness of the emission is usually identified with the
maximum virtuality that the emitted gluon can have without perturbing
significantly the collinear singularity in Eq. (\ref{eq:massiveden}). In fact,
the argument of the running strong coupling constant is set to this
virtuality. If we assume that $k$ develops a positive virtuality, with $k^0$
being replaced by $\tilde{k}^0 = \sqrt{k^2 + k_0^2}$, the variation of the
denominator of Eq.~(\ref{eq:massiveden}) is
\begin{equation}
  \delta \left[ (k + p)^2 - m^2 \right] = ( \tilde{k}_0 + p_0)^2 - (k_0 +
  p_0)^2 = k^2 + 2 p_0 ( \tilde{k}_0 - k_0) \approx k^2 + \frac{p^0}{k^0} k^2 \, ,
  \label{eq:deltaden}
\end{equation}
where we have assumed $k^2 \ll k_0^2$. The second term on the r.h.s.
of Eq.~(\ref{eq:deltaden}) is dominant in the soft region, so we require
\begin{equation}
  \frac{p^0}{k^0} k^2 < (k + p)^2 - m^2,\;\; \tmop{or}\;\;
 k^2 < \frac{k^0}{p^0} 2 k 
  \cdot p \, ,
\end{equation}
in order for the virtuality of $k$ not to alter significantly the collinear
region. Thus, we can use as hardness definition
\begin{equation}
  K_{\perp}^2 = \frac{k^0}{p^0} 2 k \cdot p \, ,
\end{equation}
that in the massless limit becomes
\begin{equation}
  \frac{k^0}{p^0} 2 k \cdot p = 2 (k^0)^2 (1 - \cos \theta) \, ,
\end{equation}
which corresponds to the \tmtexttt{POWHEG} \tmtexttt{BOX} hardness definition
for final state singularities in the massless limit.

We now examine the structure of the $d \Phi_{\tmop{rad}}$ phase space with a
constraint on $K_{\perp}^2$. Both $k \cdot p$ and $p^0$ have simple
expressions in terms of the $\bar{\Phi}_n, \Phi_{\tmop{rad}}$ variables. We
have (see Eq.~(\ref{eq:zandcsi}))
\begin{equation}
  p^0 = \bar{p}^0 - z \xi \frac{q}{2} \, ,
\end{equation}
while $p \cdot k$ is obtained from
\begin{equation}
  (q - k_{\tmop{rec}})^2 = (p + k)^2 = m^2 + 2 p \cdot k \, ,
\end{equation}
which yields
\begin{equation}
  q^2 + M_{\tmop{rec}}^2 - 2 qk_{\tmop{rec}}^0 = m^2 + 2 p \cdot k \, ,
\end{equation}
that using the second formula in Eq.~(\ref{eq:z12kbrec}) provides
\begin{eqnarray}
  p \cdot k & = & q ( \bar{k}_{\tmop{rec}} - k_{\tmop{rec}}) \, . 
\end{eqnarray}
Now
\begin{equation}
  k_{\tmop{rec}} = q - k^0 - p^0 = q - \left( \xi \frac{q}{2} + \bar{p}^0 - z
  \xi \frac{q}{2} \right) \, ,
\end{equation}
while $\bar{k}_{\tmop{rec}}$ is obtained by setting $\xi = 0$ in the above
expression. Thus
\begin{equation}
  p \cdot k = q ( \bar{k}_{\tmop{rec}} - k_{\tmop{rec}}) = \xi \frac{q^2}{2}
  (1 - z) \, . \label{eq:pdotk}
\end{equation}
Summarizing
\begin{equation}
  K_{\perp}^2 = \frac{\xi^2 q^3 \left( 1 - z \right)}{2 \bar{p}^0 - z \xi q} \, .
\end{equation}
This expression seems to be still a bit complex. On the other hand, a theta
function
\begin{equation}
  \theta (K_{\perp}^2 - t) = \theta \left( \xi^2 q^3 - 2 tp_{\max}^0 - (\xi
  q^2 - t) z \xi q \right) \, ,
\end{equation}
is easily solved in $z$, so that keeping the full expression is in fact a
viable option.

It is useful to have a view of the constant $K_T^2$ curves in the Dalitz
plane. We see that these curves are all decreasing as a function of $k^0$, and
they cross the boundary of the Dalitz region just once. It will be convenient,
in our case, to consider the extended region
\begin{equation}
  z_2 < z < z_1 \, , \, \, \, 0 < \xi < \xi_{\max} \, , \label{eq:extendedregion}
\end{equation}
where
\begin{equation}
  \xi_{\max} = 1 - \frac{(m + m_{\tmop{rec}})^2}{q^2} \, .
\end{equation}
This extended region can be used to generate radiation according to the upper
bound on $R / B$, where phase space points generated outside the real Dalitz
region will be vetoed. Now, we can prove that, in the extended region
as in Eq.~(\ref{eq:extendedregion}), on the lines of constant $K_{\perp}^2$, $\xi$ is
monotonically increasing as $z$ increases. In fact, solving for $\xi$ at fixed
$z$, we find one positive solution
\begin{equation}
  \xi = \frac{\sqrt{K_{\perp}^2 \left( K_{\perp}^2 z^2 + 8 \bar{p}^0 q (1 - z)
  \right)} - K_{\perp}^2 z}{2 q^2 (1 - z)} \, . \label{eq:xivsk2}
\end{equation}
It can be easily checked that the above function has positive derivative with
respect to $z$ for $K_{\perp}^2 < 2 \bar{p}^0 q$, and negative derivative for
$K_{\perp}^2 > 2 \bar{p}^0 q$. For the special value $K_{\perp}^2 = 2
\bar{p}^0 q$, the above equation yields $\xi = 2 \bar{p}^0 / q$, independent
on $z$. On the other hand, this value is above $\xi_{\max}$
\begin{equation}
  2 \bar{p}^0 / q = \frac{q^2 + m^2 - M_{\tmop{rec}}^2}{q^2} > 1 - \frac{(m +
  M_{\tmop{rec}})^2}{q^2} \, ,
\end{equation}
and therefore is never reached. Furthermore, Eq. (\ref{eq:xivsk2}) is an
increasing function of $K_{\perp}^2$, so, for $K_{\perp}^2 > 2 \bar{p}^0 q$ we
have $\xi > \xi_{\max}$. Therefore, we always have $K_{\perp}^2 < 2 \bar{p}^0
q$ in the region of Eq.~(\ref{eq:extendedregion}), and the function of Eq.~(\ref{eq:xivsk2}) has
positive derivative in $z$. Thus, the minimum value of $\xi$ at given
$K_{\perp}^2$ in the region of Eq.~(\ref{eq:extendedregion}) is at the smallest value of
$z$, {\it i.e.} $z_2$
\begin{equation}
  \xi_{\min} (K_{\perp}^2) = \frac{\sqrt{K_{\perp}^2 \left( K_{\perp}^2 z_2^2
  + 8 \bar{p}^0 q (1 - z_2) \right)} - K_{\perp}^2 z_2}{2 q^2 (1 - z_2)} \, .
\end{equation}
If $\xi_{\min} (K_{\perp}^2) > \xi_{\max}$, this $K_{\perp}$ value is
forbidden. The maximum value of $K_{\perp}$ is easily obtained, since it
corresponds to $\xi_{\min} (K_{\perp}^2) = \xi_{\max}$ and $z = z_2$
\[ t_{\max} \equiv \max (K_{\perp}^2) = \frac{\xi_{\max}^2 q^3 \left( 1 - z_2
   \right)}{2 \bar{p}^0 - z_2 \xi_{\max} q} \, . \]
The integral of a phase space function $f$ with a $K_{\perp}^2 > t$ cut,
assuming $t < t_{\max}$, is thus given by
\begin{eqnarray}
 \!\!\!\!\!\!\! I_f (t) = \!\! \int_0^{\xi_{\max}} d \xi \int_{z_2}^{z_1} dz \theta \left(
  K_{\perp}^2 - t \right) f (\xi, z) & = & \!\!\! \int_{\xi_{\min} (t)}^{\xi_{\max}}
  d \xi \int^{\min (z_1, z_{\max} (t, \xi))}_{z_2} dzf (\xi, z), 
\end{eqnarray}
where
\begin{equation}
  z_{\max} (t, \xi) = \frac{\xi^2 q^3 - 2 t \bar{p}^0}{\xi q (\xi q^2 - t)} \, ,
  \label{eq:zmaxdef}
\end{equation}
{\it i.e.} is the value of $z$ such that $K_{\perp}^2 = t$.

It is now convenient to further break this integral. Since $z_{\max}$ is an
increasing function of $\xi$, there is going to be a value $\xi_1$ such that
for $\xi < \xi_1$ we have $z_{\max} (t, \xi) < z_1$. It is obviously given by
\begin{equation}
  \xi_1 (t) = \frac{\sqrt{t \left( tz_1^2 + 8 \bar{p}^0 q (1 - z_1) \right)} -
  tz_1}{2 q^2 (1 - z_1)} \, .
\end{equation}
So, we can further break the phase space as
\begin{equation}
  I_f (t) = \int_{\xi_{\min} (t)}^{\min (\xi_1 (t), \xi_{\max})} d \xi
  \int^{z_{\max} (t, \xi))}_{z_2} dzf (\xi, z) + \theta (\xi_{\max} - \xi_1
  (t)) \int_{\xi_1 (t)}^{\xi_{\max}} d \xi \int^{z_1}_{z_2} dzf (\xi, z) \, .
\end{equation}

\subsection{Approximation to $R / B$}

The eikonal approximation to the real amplitude $R$ yields
\begin{equation}
  \mathcal{A}_R = \mathcal{A}_B \left( \frac{p^{\mu}}{p \cdot k} -
  \frac{r^{\mu}}{r \cdot k} \ldots \right) \, ,
\end{equation}
and squaring, and including the $- g_{\mu \nu}$ from the spin projection, we
get
\begin{equation}
  \mathcal{A}_R^2 = \mathcal{A}_B^2 \left( - \frac{m^2}{(p \cdot k)^2} +
  \frac{2 p \cdot r}{p \cdot kr \cdot k} \ldots \right) \, .
\end{equation}
The $m^2$ term cannot make the cross section negative, and thus is bounded
by the other terms. Separating the region of $k$ collinear to $p$ in the usual
way
\begin{equation}
  \frac{p \cdot r}{p \cdot kr \cdot k} \approx \frac{1}{p \cdot k} 
  \frac{p^0}{k^0} \propto \frac{1}{p \cdot k}  \frac{q}{2 k^0} =
  \frac{1}{\xi^2 q^2 \left( 1 - z \right)} \text{} \frac{}{} \text{} \, ,
  \label{eq:robapprox}
\end{equation}
where we have used Eq.~(\ref{eq:pdotk}). The form on the r.h.s. of
Eq.~(\ref{eq:robapprox}) is meant to capture the singular behaviour of the
amplitude, so that it may serve as an upper bound of it. The jacobian for real
radiation is
\begin{equation}
  \frac{1}{(2 \pi)^3} \frac{q^2}{4 \bar{p}} k^0 d \xi dzd \phi \, .
  \label{eq:phspacer}
\end{equation}
We can thus assume as an upper bound to the $R / B$ expression the form
(\ref{eq:robapprox}), extended to the whole region of Eq.~(\ref{eq:extendedregion}) 
$0 < \xi < 1$. It is useful to impose a further restriction on the region (\ref{eq:extendedregion}), 
namely
\begin{equation}
  K_{\perp}^2 \leqslant q^2 \, .
\end{equation}
This further constraint includes the physical region, because
\begin{equation}
  K_T^2 = \frac{k^0}{p^0} 2 k \cdot p \leqslant \frac{k^0}{p^0} 4 k^0 p^0 = 4
  (k^0)^2 = q^2 \xi^2 \, .
\end{equation}
Summarizing, by multiplying Eq.~(\ref{eq:robapprox}) by the phase space 
of Eq.~(\ref{eq:phspacer}), our upper bounding function must have the form
\begin{equation}
  \frac{q}{\bar{p}} \frac{1}{\xi \left( 1 - z \right)} \text{} \frac{}{}
  \text{} d \xi dzd \phi \, .
\end{equation}

\subsection{Integration}

We now evaluate the integral of the upper bounding function. We get
\begin{eqnarray}
  I (t) & = & \int_{\xi_{\min} (t)}^{\min (\xi_1 (t), \xi_{\max})} d \xi
  \int^{z_{\max} (t, \xi)}_{z_2} \frac{q}{\bar{p}} \frac{dz}{\xi (1 - z)} +
  \theta (\xi_{\max} - \xi_1 (t)) \int_{\xi_1 (t)}^{\xi_{\max}} d \xi
  \int^{z_1}_{z_2} dz \frac{q}{\bar{p}} \frac{1}{\xi (1 - z)} \nonumber\\
  & \! = \! & \! \frac{q}{\bar{p}} \int_{\xi_{\min} (t)}^{\min (\xi_1 (t),
  \xi_{\max})} \frac{d \xi}{\xi} \log \frac{1 - z_2}{1 - z_{\max} (t, \xi)} +
  \frac{q}{\bar{p}} \theta (\xi_{\max} - \xi_1 (t)) \log
  \frac{\xi_{\max}}{\xi_1 (t)} \log \frac{1 - z_2}{1 - z_1} . \label{eq:inti1}
\end{eqnarray}
From Eq. (\ref{eq:zmaxdef}) we have
\begin{equation}
  1 - z_{\max} (t, \xi) = \frac{2 t \bar{p}^0 - \xi qt}{\xi q (\xi q^2 - t)} \, .
\end{equation}
We can break up the logarithm in the first integral of Eq.~(\ref{eq:inti1}) as
\begin{equation}
\log \frac{(1 - z_2) \xi q (\xi q^2 - t)}{t (2 \bar{p}^0 - \xi q)} = \log
   \left[ (1 - z_2) \frac{q}{t} \right] + \log \xi + \log (\xi q^2 - t) - \log
   (2 \bar{p}^0 - \xi q) \, .
\end{equation}
Thus all the integrals have the form
\begin{equation}
  \int \frac{d \xi}{\xi} \log (a + b \xi) \, ,
\end{equation}
with the argument of the logarithm being positive in the integration range. Thus it
can be expressed in terms of a dilogarithm
\begin{equation}
  \int \frac{d \xi}{\xi} \log (a + b \xi) = G (a, b, \xi) + \tmop{constant} \, ,
\end{equation}
with
\begin{equation}
  G (a, b, \xi) \equiv \log (a + b \xi) \log \left( 1 - \frac{a + b \xi}{a}
  \right) + \tmop{Li}_2 \left( \frac{a + b \xi}{a} \right) \, \, \tmop{for} \, \, a < 0 \, ,
\end{equation}
\begin{equation}
  G (a, b, \xi) \equiv \log | \frac{b \xi}{a} | \log a - \tmop{Li}_2 \left[ -
  \frac{b \xi}{a} \right] + \frac{\pi^6}{6} \,\, \tmop{for} \,\, a > 0 \,.
\end{equation}
We thus get
\begin{eqnarray}
  I (t) & = & \frac{q}{\bar{p}} \left[ \log \xi \log \left[ (1 - z_2)
  \frac{q}{t} \right] + \frac{1}{2} \log^2 \xi + G (- t, q^2, \xi) - G (2
  \bar{p}^0, - q, \xi) \right]^{\min (\xi_1 (t), \xi_{\max})}_{\xi_{\min}}
  \nonumber\\
  & + \frac{q}{\bar{p}} & \theta (\xi_{\max} - \xi_1 (t)) \log
  \frac{\xi_{\max}}{\xi_1 (t)} \log \frac{1 - z_2}{1 - z_1} \,. 
\end{eqnarray}

\subsection{Generation of $z$ and $\xi$ at fixed $t$}

We now see how $z$ and $\xi$ can be generated once $t$ has been found. They
are distributed according to
\begin{equation}
  \frac{d \xi dz}{\xi (1 - z)} \delta \left( \frac{\xi^2 q^3 (1 - z)}{2
  \bar{p}^0 - z \xi q} - t \right) \,. \label{eq:csizdist}
\end{equation}
Performing first the $z$ integration, we get
\begin{equation}
  d \xi \frac{q^2}{t (\xi q^2 - t)} \, .
\end{equation}
Thus we first generate $\xi$ uniformly in $\log (\xi q^2 - t)$. The extremes
for $\xi$ are given by $\xi_{\min} (t)$ and $\xi_m (t) = \min (\xi_{\max},
\xi_1 (t))$
\begin{equation}
  \xi = \left\{ \exp \left[ \log (\xi_{\min} (t) q^2 - t) + r \log \frac{\xi_m
  (t) q^2 - t}{\xi_{\min} (t) q^2 - t} \right] + t \right\} / q^2 \, ,
\end{equation}
where $0 < r < 1$ is a uniform random number. The value of $z$ is then
obtained by solving the $\delta$ function in Eq.~(\ref{eq:csizdist}).

\bibliographystyle{JHEP}
\bibliography{POWHEGEW-revised}

\end{document}